\documentstyle[11pt]{article}

\newlength{\minitwocolumn}
\catcode`\@=11
\def\relaxnext@{\let\next\relax}
\font\tenmsy=msym10 scaled\magstep1
\font\sevenmsy=msym7 scaled\magstep1
\font\fivemsy=msym5  scaled\magstep1
\newfam\msyfam
\textfont\msyfam=\tenmsy
\scriptfont\msyfam=\sevenmsy
\scriptscriptfont\msyfam=\fivemsy
\font\teneuf=eufm10 scaled\magstep1
\font\seveneuf=eufm7 scaled\magstep1
\font\fiveeuf=eufm5 scaled\magstep1
\newfam\euffam
\textfont\euffam=\teneuf
\scriptfont\euffam=\seveneuf
\scriptscriptfont\euffam=\fiveeuf
\def\frak{\relaxnext@\ifmmode\let\next\frak@\else
 \def\next{\Err@{Use \string\frak\space only in math mode}}\fi\next}
\def\goth{\relaxnext@\ifmmode\let\next\frak@\else
 \def\next{\Err@{Use \string\goth\space only in math mode}}\fi\next}
\def\frak@#1{{\frak@@{#1}}}
\def\frak@@#1{\noaccents@\fam\euffam#1}
\def\Bbb{\relaxnext@\ifmmode\let\next\Bbb@\else
 \def\next{\Err@{Use \string\Bbb\space only in math mode}}\fi\next}
\def\Bbb@#1{{\Bbb@@{#1}}}
\def\Bbb@@#1{\noaccents@\fam\msyfam#1}
\def\accentfam@{7}
\def\noaccents@{\def\accentfam@{0}}
\catcode`\@=\active

\def\eq#1\endeq{\begin{eqnarray}#1\end{eqnarray}}
\def\eqn#1\endeqn{\begin{eqnarray*}#1\end{eqnarray*}}

\makeatletter
\@addtoreset{equation}{section}
\makeatother

\newtheorem{thm}{Theorem}[section]
\newtheorem{prop}[thm]{Proposition}
\newtheorem{lem}[thm]{Lemma}
\newtheorem{cor}[thm]{Corollary}

\begin{document}
\begin{flushright}
RIMS-1075 \\
to appear in J. Stat. Phys.
\end{flushright}
\vspace{24pt}
\begin{center}
\begin{Large}
{\bf Ground state Correlation functions for an impenetrable Bose gas \\
with Neumann or Dirichlet boundary conditions.}
\end{Large}

\vspace{30pt}
By
\vspace{4pt}

\vspace{2pt}
Takeo Kojima\raisebox{2mm}{$\star$}
\vspace{6pt}

{\it Research Institute for Mathematical Sciences,
     Kyoto University, Kyoto 606, Japan}
\vspace{2pt}

{\rm e-mail address : kojima@kurims.kyoto-u.ac.jp}

\vspace{60pt}

\underline{Abstract}

\end{center}

We study density correlation functions for an impenetrable Bose gas
in a finite box,
with Neumann or Dirichlet boundary conditions in the ground state.
We derive the Fredholm minor determinant formulas for
the correlation functions.
In the thermodynamic limit, we express the correlation functions
in terms of solutions of non-linear differential equations
which were introduced by Jimbo, Miwa, M\^ori and Sato
as a generalization of the fifth Painlev\'e equations.

\vspace{25pt}

\vspace{15pt}

\vfill
\hrule

\vskip 3mm
\begin{footnotesize}

\noindent\raisebox{2mm}{$\star$}
Research Fellow of the Japan Society for the Promotion
of Science.
\end{footnotesize}
\newpage
\section{Introduction}
In the standard treatment of quantum integrable systems,
one starts with a finite box and imposes periodic boundary conditions,
in order to ensure integrability.
Recently, there has been increasing interest in exploring other possible
boundary conditions compatible with integrability.

With non-periodic boundary conditions,
the works on the Ising model are among the earliest.
By combinatorial arguments, 
McCoy and Wu \cite{M.W.} studied the two-dimensional Ising model with a
general boundary.
They calculated the spin-spin correlation functions of
two spins in the boundary row.
Using fermions,
Bariev \cite{Bariev1} studied the two-dimensional Ising model with a Dirichlet
boundary. He calculated the local magnetization
and derived the third Painlev\'e differential equations
in the scaling limit.
Bariev \cite{Bariev2} generalized his calculation to 
a general boundary case.
In the Neumann boundary case, he
also derived the third Painlev\'e differential equations
in the scaling limit.
Sklyanin \cite{Sklyanin} began a systematic approach 
to open boundary problems,
so-called open boundary Bethe Ansatz.
Jimbo et al \cite{J.K.K.K.M.} calculated
correlation functions of local operators
for antiferromagnetic XXZ chains with a general boundary,
using Sklyanin's algebraic framework and the
representation theory of quantum affine algebras.

Sklyanin \cite{Sklyanin} explained the integrability
of the open boundary
impenetrable bose gas model, using boundary
Yang Baxter equations.
In this paper, we will study field correlation functions
(density matrix) for an impenetrable bose gas with Neumann
or Dirichlet boundary conditions.
Schultz \cite{Schultz} studied field correlation functions 
for an impenetrable bose gas
with periodic boundary conditions.
He discretized the second quantized Hamiltonian
and found that the discretized Hamiltonian was the
isotropic XY model Hamiltonian.
He diagonalized the discretized Hamiltonian by
introducing fermion operators.
Using the $N$ particle ground state eigenvector for
the discretized Hamiltonian, Schultz derived
an explicit formula of correlation functions
for an impenetrable bose gas in the continuum limit.
Lenard \cite{Lenard} pointed out
that Schultz's formula could be written by
Fredholm minor determinants.
Therefore this formula
is called Schultz-Lenard formula.
In this paper, we will derive Schultz-Lenard type formula
for Neumann or Dirichlet boundary condition.
Following Schltz, we employ two devices.
We consider the $N$ particle ground state of the discretized Hamiltonian.
We then fermionize the discretized $N$ particle system by using the
Jordan-Wigner transformation.
In the continuum limit, we derive the Fredholm minor
determinant formula for correlation function,
which has the integral kernel :
\begin{eqnarray}
\frac{\pi}{2L}\left\{\frac{\sin\frac{2N+1}{2L}\pi(x-x')}
{\sin \frac{1}{2L}\pi(x-x')}+\varepsilon
\frac{\sin\frac{2N+1}{2L}\pi(x+x')}
{\sin \frac{1}{2L}\pi(x+x')}\right\},
\end{eqnarray}
\begin{eqnarray}
( L: box~size,~ N: the~number~of~particles, ~\varepsilon=+:Neumann, 
~\varepsilon=-:Dirichlet.)\nonumber
\end{eqnarray}
Jimbo, Miwa, M\^ori and Sato
\cite{J.M.M.S.}
developed the deformation theory for Fredholm integral equation 
of the second kind
with the special kernel $\frac{\sin(x-x')}{x-x'}$.
They introduced a system of nonlinear
partial differential equation,
which becomes the fifth Painlev\'e in the simplest case.
They showed that the correlation functions
without boundaries was
the $\tau$-function of their generalization of the fifth
Painlev\'e equations.
In this paper,
we express the correlation functions for Neumann
or Dirichlet boundaries in terms of solutions
of Jimbo, Miwa, M\^ori and Sato's generalization of
the fifth Painlev\'e equations,
hereafter refered to as the JMMS equations.
In the thermodynamic limit
$(N, L \to \infty, \frac{N}{L}:fixed)$,
we reduce the differential equations
for correlation functions with
Neumann or Dirichlet boundaries to
that without boundaries,
using the reflection relation between
two integral kernels
$\frac{\sin(x-x')}{x-x'}+\varepsilon
\frac{\sin(x+x')}{x+x'}$ and
$\frac{\sin(x-x')}{x-x'}$.
The two point correlation function with Neumann boundary
is described by the equation (\ref{tau}) and (\ref{Hamil.0}).
In the case with boundary,
the differential equation for the two point correlation function
cannot be described by an ordinary differential equation.
We need three variable case of the JMMS equations.

Physically, the long distance asymptotics of the correlation function
are interesting. 
The long distance asymptotics of the ordinary differential
Painleve V is known. But, for many variable case,
the asymptotics of the JMMS equations
are not known. Therefore we cannot describe the long distance asymptotics
of the correlation functions with boundary in this paper.
To evaluate the asymptotics of the solution of 
the JMMS equation is our future problem.

Now a few words about the organization of the paper.
In section 2, we state the problem and summarize the
main results.
In section 3, we derive an explicit formula
for the correlation functions in a finite box.
In section 4, we write down the differential equations
for the correlation functions in the thermodynamic limit.

\section{Formulation and results}
The purpose of this section is to formulate the problem and 
summarize the main results.
The quantum mechanics problem we shall study
is defined by the following four conditions.
Let $N \in {\bf{N}}, ({\bf{N}}\geq 2), L \in {\bf{R}}
, \vartheta_0, \vartheta_L \in {\bf{R}}.$
\begin{enumerate}
\item The wave function $\psi_{N,L}=\psi_{N,L}
(x_1,\cdots ,x_N \vert \vartheta_0, \vartheta_L)$
satisfies the free-particle Schr\"odinger equation for the motion of
$N$ particles in one dimension 
$(0 \leq (x_i \neq x_j) \leq L)$.
Here the variables $x_1, \cdots, x_N$
stand for the coordinates of the particles.
\item The wave function $\psi_{N,L}$ is symmetric with respect to 
the coordinates.
\begin{eqnarray}
\psi_{N,L}
(x_1,\cdots ,x_N \vert \vartheta_0, \vartheta_L)
=
\psi_{N,L}
(x_{\sigma(1)},\cdots ,x_{\sigma(N)} \vert \vartheta_0, \vartheta_L)
,~~~(\sigma \in S_N).
\end{eqnarray}
\item The wave function satisfies the open boundary conditions
in a box $0 \leq x_j \leq L, ~(j=1,\cdots, N)$
\begin{eqnarray}
\left.\left(\frac{\partial}{\partial x_j}-\vartheta_0 \right)
\psi_{N,L}(x_1,\cdots ,x_N \vert \vartheta_0, \vartheta_L)~
\right|_{x_j=0}=0,~~(j=1, \cdots, N),\\
\left.\left(\frac{\partial}{\partial x_j}+\vartheta_L \right)
\psi_{N,L}(x_1,\cdots ,x_N \vert \vartheta_0, \vartheta_L)~
\right|_{x_j=L}=0,~~(j=1, \cdots, N).
\end{eqnarray}
\item The wave function $\psi_{N,L}$ vanishes whenever two particle
coordinates coincide. 
\begin{eqnarray}
\psi_{N,L}
(x_1,\cdots ,x_i, \cdots, x_j,\cdots ,x_N \vert \vartheta_0, \vartheta_L)
=0,~~{\rm for}~~~
x_i=x_j.
\end{eqnarray}
\end{enumerate}

In this paper we shall be concerned with the ground state.
The wave function is given by
\begin{eqnarray}
\psi_{N,L}(x_1,\cdots, x_N \vert \vartheta_0, \vartheta_L)
=\frac{1}{\sqrt{V_{N,L}(\vartheta_0, \vartheta_L)}}
\left|
\det_{1 \leq j,k \leq N}
\left( \lambda_j \cos \left(\lambda_j x_k \right)
+\vartheta_0 \sin \left( \lambda_j x_k \right)\right)
\right|.
\end{eqnarray}
Here the momenta
$ 0 <\lambda_1 < \cdots < \lambda_N $
are determined from the boundary condition for
$\psi_{N,L}$ which amounts to the equations
\begin{eqnarray}
2L\lambda_j+\theta_{\vartheta_0}(\lambda_j)
+\theta_{\vartheta_L}(\lambda_j)=2 \pi j,~~(j=1, 
\cdots , N),\label{BAE}
\end{eqnarray}
where we set
$\theta_d(\lambda)
=i{\rm log}
\left(
\frac{id+\lambda}
{id-\lambda}\right)$. 
We take the branch
$-\pi <\theta_d(\lambda) <\pi,~ (d \geq 0).$
Here $V_{N,L}(\vartheta_0, \vartheta_L)$
is a normalization factor defined by
\begin{eqnarray}
V_{N,L}(\vartheta_0, \vartheta_L)
=\frac{N!}{2^{2N}}
\det_{1 \leq j,k \leq N}
\left(
\sum_{\varepsilon, \varepsilon'=\pm}
(\lambda_j+\varepsilon \vartheta_0)
(\lambda_k+\varepsilon' \vartheta_0)
\int_0^Le^{i(\varepsilon \lambda_j
-\varepsilon'\lambda_k)y} dy \right).
\end{eqnarray}
The wave function is not translationally invariant and has the normalization,
\begin{eqnarray}
\int_0^L\cdots \int_0^L
dy_1 \cdots dy_N 
\psi_{N,L}(y_1, \cdots, y_N \vert \vartheta_0, \vartheta_L)^2=1.
\end{eqnarray}
The following equation holds for any parameter $\lambda$,
\begin{eqnarray}
\left.\left(\frac{\partial}{\partial x}-\vartheta_0\right)
(\lambda \cos (\lambda x)+\vartheta_0 \sin (\lambda x))\right|
_{x=0}=0.\label{wave1}
\end{eqnarray}
The following equivalent relation holds,
\begin{eqnarray}
\left.\left(\frac{\partial}{\partial x}+\vartheta_L\right)
(\lambda \cos (\lambda x)+\vartheta_0 \sin (\lambda x))\right|
_{x=L}=0 ~\Leftrightarrow~
e^{2 i L\lambda}
=\frac{(\lambda+i \vartheta_L)(\lambda+i \vartheta_0)}
{(\lambda-i \vartheta_L)(\lambda-i \vartheta_0)}.
\label{wave2}
\end{eqnarray}
>From (\ref{wave1}), (\ref{wave2})
and Girardeau's observation
on fermions and impenetrable bosons correspondence in one dimension
\cite{Gir},
we can show that the wave function $\psi_{N,L}$
satisfies the above four conditions.
We shall be interested in the field correlation functions
(density matrix) given by
\begin{eqnarray}
&&\rho_{n,N,L}(x_1,\cdots, x_n \vert x_1',\cdots ,x_n'
\vert \vartheta_0, \vartheta_L)
\nonumber \\
&=&\frac{(n+N)!}{N!}
   \int_{0}^{L} \cdots \int_{0}^{L} dy_{n+1}\cdots dy_{n+N}
   \psi_{n+N, L} (x_1,\cdots,x_n,y_{n+1},\cdots,y_{n+N}
\vert \vartheta_0, \vartheta_L)
    \nonumber \\
&&~~~~~~~~~\times
   \psi_{n+N, L} (x_{1}',\cdots,x_{n}',y_{n+1},\cdots,y_{n+N}
\vert \vartheta_0, \vartheta_L).
\end{eqnarray}
In this paper, following \cite{Schultz}, we reduce
our problem to that of discrete $M$ intervals.
Set $\epsilon=\frac{L}{M+1}$.
Let
$\vert v_1\rangle
=\left(\begin{array}{c}
1\\
0
\end{array}\right),~
\vert v_2\rangle
=\left(\begin{array}{c}
0\\
1
\end{array}\right)$
be the standard basis of $V={\bf C}^{2}$.
Let
$\langle v_i \vert ,~(i=1,2)$ be the dual basis given by
$\langle v_i \vert v_j \rangle =\delta_{i,j} ,~(i,j=1,2)$.
The action of ${\it O}\in End\left( {\bf C}^2 \right)$
on $\langle v_i \vert ,~(i=1,2)$ is defined by
$\left( \langle v_i \vert {\it O}\right) \vert v_j \rangle
:=\langle v_i \vert \left( {\it O} \vert v_j \rangle \right),~(j=1,2)$.
Set $\vert \Omega_0 \rangle =\vert v_1 \rangle^{\otimes M}$
and $\langle\Omega_0 \vert
=\langle v_1 \vert^{\otimes M}$.
Set
\begin{eqnarray}
\phi^{+}=
\left(
\begin{array}{cc}
0& 0 \\
1& 0
\end{array}\right),~~~
\phi=
\left(
\begin{array}{cc}
0& 1 \\
0& 0
\end{array}\right),~~
\sigma^{z}=
\left(
\begin{array}{cc}
1& 0 \\
0& -1
\end{array}\right).
\end{eqnarray}
Following the usual convention, we let
$\phi_j^+, \phi_j, \sigma^z_j$
signify the operators acting on the $j-$th tensor
component of $V^{\otimes M}$.
Introduce fermion operators 
$\psi^+_m, \psi_m $ by the Jordan-Wigner transformation
\begin{eqnarray}
\psi_m^+=\sigma^z_1 \cdots \sigma_{m-1}^z \phi^+_m,~~ 
\psi_m=\sigma^z_1 \cdots \sigma_{m-1}^z \phi_m,~
(m=1,\cdots ,M).
\end{eqnarray}
The fermion operators have the anti-commutation relations
\begin{eqnarray}
\{\psi_m^+, \psi_n\}=\delta_{m,n},~~ 
\{\psi_m, \psi_n\}=\{\psi_m^+, \psi_n^+\}=0.
\end{eqnarray}
Here we use the notation
$\{a,b\}=ab+ba.$
Set 
\begin{eqnarray}
\vert \Omega_{N,M}(\vartheta_0, \vartheta_L)\rangle
&=&\sqrt{\frac{1}{N!}}\sum_{m_1,\cdots,m_N=1}^M
\psi_{N,L}(\epsilon m_1, \cdots ,\epsilon m_N \vert 
\vartheta_0, \vartheta_L)
\phi_{m_1}^+ \cdots \phi_{m_N}^+
\vert \Omega_0 \rangle  \\
&=&\frac{1}{\sqrt{N!~ V_{N,L}(\vartheta_0, \vartheta_L)}}
\prod_{j=1}^N \sum_{m_j=1}^M
\left(\lambda_j \cos (\epsilon m_j \lambda_j)
-\vartheta_0 \sin (\epsilon m_j \lambda_j) \right)
\psi_{m_1}^+ \cdots \psi_{m_N}^+ \vert \Omega_0 \rangle,\nonumber \\
\langle \Omega_{N,M}(\vartheta_0, \vartheta_L)\vert
&=&\sqrt{\frac{1}{N!}}\sum_{m_1,\cdots,m_N=1}^M
\psi_{N,L}(\epsilon m_1, \cdots ,\epsilon m_N \vert
\vartheta_0, \vartheta_L)\langle \Omega_0 \vert
\phi_{m_1} \cdots \phi_{m_N} \\
&=&\frac{1}{\sqrt{N!~ V_{N,L}(\vartheta_0, \vartheta_L)}}
\prod_{j=1}^N \sum_{m_j=1}^M
\left(\lambda_j \cos (\epsilon m_j \lambda_j)
-\vartheta_0 \sin (\epsilon m_j \lambda_j) \right)
\langle \Omega_0 \vert \psi_{m_1}\cdots \psi_{m_N}.\nonumber
\end{eqnarray}
Using the above vectors, we can calculate
correlation functions in the continuum limit as
\begin{eqnarray}
&&\rho_{n,N,L}(x_1,\cdots,x_n \vert x_1',\cdots,x_n'
\vert \vartheta_0, \vartheta_L)
\nonumber\\
&=&
\lim_{M \to \infty}
\left(\frac{L}{M}\right)^N
\langle \Omega_{n+N,M}(\vartheta_0, \vartheta_L) \vert
\phi_{s_1}\phi_{s_2}\cdots
\phi_{s_n}
\phi^+_{t_1}\phi^+_{t_2}\cdots
\phi^+_{t_n}
\vert \Omega_{n+N,M}(\vartheta_0, \vartheta_L) \rangle,
\label{continuum}
\end{eqnarray}
where we take the limit $M \to \infty$ in such a way that
$\epsilon s_j \to x_j, \epsilon t_j \to x_j' , (L: fixed)$.
The equation (\ref{continuum})
follows from (\ref{con1}) and (\ref{con2}).
\begin{eqnarray}
&&n^2(_{N+n}C_n)^2 N!~~
\prod_{j=1}^N \sum_{m_j=1 \atop{m_j \neq t_1,\cdots ,t_n,
s_1, \cdots ,s_n}}^M 
\langle \Omega_0 \vert
\phi_{m_1}\cdots \phi_{m_N}\phi_{m_1}^{+}\cdots \phi_{m_N}^{+}
\vert \Omega_0 \rangle 
\nonumber \\
&\times &
\psi_{n+N,L}(
\epsilon t_1\cdots \epsilon t_n, \epsilon m_1\cdots \epsilon m_N
\vert \vartheta_0, \vartheta_L)~
\psi_{n+N,L}(
\epsilon t_1\cdots \epsilon t_n, \epsilon m_1\cdots \epsilon m_N
\vert \vartheta_0, \vartheta_L)
\nonumber \\
&=&
\prod_{i=1}^{N+n}\sum_{m_i=1}^M \prod_{j=1}^{N+n}\sum_{l_j=1}^M
\langle \Omega_0 \vert
\phi_{m_1}\cdots \phi_{m_{n+N}}
\phi_{s_1}\cdots \phi_{s_n}
\phi_{t_1}\cdots \phi_{t_n}
\phi_{l_1}\cdots \phi_{l_{n+N}}
\vert \Omega_0 \rangle
\nonumber\\
&\times&\psi_{n+N,L}(\epsilon m_1 \cdots \epsilon m_{n+N}
\vert \vartheta_0, \vartheta_L)~
\psi_{n+N,L}(\epsilon l_1 \cdots \epsilon l_{n+N}
\vert \vartheta_0, \vartheta_L),
\label{con1}
\end{eqnarray}
\begin{eqnarray}
\langle \Omega_0 \vert
\phi_{m_1}\cdots \phi_{m_N}\phi_{m_1}^{+}\cdots \phi_{m_N}^{+}
\vert \Omega_0 \rangle
=1.
\label{con2}
\end{eqnarray}
This formula (\ref{continuum})
is our standing point.
The case $\vartheta_0, \vartheta_L= 0$
corresponds to Neumann boundary condition
and the case $\vartheta_0, \vartheta_L =\infty$ to
Dirichlet boundary condition.
In the sequel, we use the following abbreviations.
\begin{eqnarray}
\rho_{n,N,L}(x_1,\cdots, x_n \vert
x_1',\cdots, x_n' \vert +)
&=&\rho_{n,N,L}(x_1,\cdots, x_n \vert
x_1',\cdots ,x_n' \vert 0,0),\\
\rho_{n,N,L}(x_1,\cdots, x_n \vert
x_1',\cdots, x_n' \vert -)
&=&\rho_{n,N,L}(x_1,\cdots, x_n \vert 
x_1',\cdots, x_n'\vert \infty, \infty).
\end{eqnarray}
In the sequel, for simplicity, we consider two important case :
Neumann boundary conditions and Dirichlet boundary conditions.

~

{\sl Remark.}~~~~
{\sl There exists the simple relations between Neumann or Dirichlet boundaries
and periodic boundaries.
We can embed the differential equations for $n$ point correlation functions
of Neumann or Dirichlet boundaries, to the one for
$2n$ or $2n-1$ point correlation functions without boundaries.}

~

In section 3, we derive the following formula.
\begin{thm}~~~~
The correlation functions for an impenetrable bose gas
with Neumann or Dirichlet boundaries are given by
the following formulas.
\begin{eqnarray}
\rho_{n,N,L}(x_1',\cdots ,x_n' \vert x_{1}'',\cdots ,x_{n}''
\vert \varepsilon )
&=&\left(-\frac{1}{2} \right)^n
\prod_{1 \leq j<k \leq n}{\rm sgn}(x_k'-x_j')~{\rm sgn}(x_k''-x_j'')
\nonumber \\
&\times&\det
\left(
1- \frac{2}{\pi} {\hat K}_{\varepsilon,N,I_p}\left|
\begin{array}{cccc}
x_1', & x_2', & \cdots & ,x_n'\\
x_1'',& x_2'',& \cdots & ,x_n''
\end{array} \right. \right),
\end{eqnarray}
where $\varepsilon=\pm$ and $ 0 \leq x_j', x_j''\leq L,~ (j=1,\cdots,n)$.
Here $I_p$ is the union of $n$ intervals
$I_p=[x_1,x_2]\cup \cdots \cup [x_{2n-1},x_{2n}]$,
where $0\leq x_1\leq \cdots \leq x_{2n}\leq L$
is the re-ordering of
$x_1',\cdots,x_n', x_1'',\cdots,x_n''.$
The symbol $~\det
\left( 1-\lambda {\hat K}_{\varepsilon,N,I_p} \left|
\begin{array}{cccc}
x_1', & x_2', & \cdots & ,x_n'\\
x_1'',& x_2'',& \cdots & ,x_n''
\end{array} \right. \right)$
denotes the n-th Fredholm minor corresponding to
the following Fredholm type integral equation of
the second kind.
\begin{equation}
\left(\left( 1-\lambda {\hat K}_{\varepsilon,N,I_p} \right)f \right)(x)
=g(x),~~(x \in I_p).
\end{equation}
Here the integral operator ${\hat K}_{\varepsilon,N,I_p}$ is defined by
\begin{eqnarray}
\left({\hat K}_{\varepsilon,N,I_p}f\right)(x)
=\int_{I_p}
\frac{\pi}{2L}\left\{
\frac{\sin\frac{2(n+N)+1}{2L}\pi (x-y)}
     {\sin\frac{1}{2L}\pi (x-y)}
    +\varepsilon
\frac{\sin\frac{2(n+N)+1}{2L}\pi (x+y)}
     {\sin\frac{1}{2L}\pi (x+y)}\right\}f(y)~dy.
\end{eqnarray}\label{Th:FD1}
\end{thm}
Using the above Fredholm minor formulas,
we can take the thermodynamic limit for correlation functions, i.e.,
$N,L \to \infty, \frac{N}{L}=\rho_0~$: fixed.
\begin{cor}~~~~The correlation functions for an impenetrable bose gas
with Neumann or Dirichlet boundaries are given by
the following formulas in the thermodynamic limit.
\begin{eqnarray}
&&\rho_{n}(x_1',\cdots ,x_n' \vert x_1'',\cdots,x_n''\vert \varepsilon)
=\lim_{N,L \to \infty,~\frac{N}{L}=\rho_0}
\rho_{n, N, L}(x_1',\cdots ,x_n' \vert x_1'',\cdots,x_n''
\vert \varepsilon)\\
&=&\left(-\frac{1}{2} \right)^n
\prod_{1 \leq j<k \leq n}{\rm sgn}(x_k'-x_j')~{\rm sgn}(x_k''-x_j'')
\det
\left(
1- \frac{2}{\pi}{\hat K}_{\varepsilon,I_p}\left|
\begin{array}{cccc}
x_1', & x_2', & \cdots & ,x_n'\\
x_1'',& x_2'',& \cdots & ,x_n''
\end{array} \right. \right),\nonumber
\end{eqnarray}
where $0\leq x_j', x_j'' <+\infty, ~(j=1,\cdots,n)$.
The symbol $
\det
\left(
1- \lambda{\hat K}_{\varepsilon,I_p}\left|
\begin{array}{cccc}
x_1', & x_2', & \cdots & ,x_n'\\
x_1'',& x_2'',& \cdots & ,x_n''
\end{array} \right. \right),
$
represents the n-th Fredholm minor corresponding to
the following Fredholm type integral equation
of the second kind,
\begin{equation}
\left(\left( 1-\lambda{\hat K}_{\varepsilon,I_p} \right)f \right)(x)
=g(x),~~(x \in I_p),
\end{equation}
where the integral operator ${\hat K}_{\varepsilon,I_p}$ is defined by
\begin{eqnarray}
\left({\hat K}_{\varepsilon,I_p}f\right)(x)
=\int_{I_p}
\left\{
\frac{\sin\rho_0 \pi (x-y)}
     {x-y}
    +\varepsilon
\frac{\sin\rho_0 \pi (x+y)}
     {x+y} \right\}f(y)~dy.
\end{eqnarray}
\end{cor}
In the sequel, we choose such a scale
that $\pi \rho_0=1$.

In section 4, we derive the differential equations for
the correlation functions.
Jimbo, Miwa, M\^ori and Sato \cite{J.M.M.S.}
introduced the generalization of the fifth Painlev\'e equations,
hereafter refered to as the JMMS equations.
Their simplest case is exactly the fifth Painlev\'e equation.
We reduce the differential equations for
Neumann or Dirichlet boundary case
to that for without-boundary case, using
the reflection relation in lemma \ref{key}.
For $n=1$ and Dirichlet boundary case :
\begin{eqnarray}
\rho_1(0 \vert x \vert -)=0.
\end{eqnarray}
Next we explain $n=1$ and Neumann boundary case.
The differential equation for
$\rho_1(0 \vert x \vert +)$ is
described by the solutions of the Hamiltonian
equations which was introduced in \cite{J.M.M.S.}
as the special case of the generalization of the fifth Painlev\'e
equations.
We cannot describe the correlation function
$\rho_1(0 \vert x \vert +)$ in terms of the fifth
Painlev\'e ordinary differential equation.
We need the many variable case of the JMMS equations.
\begin{eqnarray}
\frac{d}{dx}{\rm log}\rho_1(0 \vert x \vert +)
=H_2(-x,0,x).\label{tau}
\end{eqnarray}
Here $H_2(a_0, a_1, a_2)$ is the coefficient
of the following Hamiltonian
\begin{eqnarray}
H&=&H_0(a_0, a_1, a_2)da_0+
H_1(a_0, a_1, a_2)da_1+H_2(a_0, a_1, a_2)da_2 \nonumber \\
&=&-\sum_{j=0,2}\frac{1}{2}(r_{+j}r_{-1}-r_{+1}r_{-j})
(\tilde{r}_{+j}r_{-1}-r_{+1}\tilde{r}_{-j})d {\rm log}(a_j-a_1)
\nonumber \\
&-&(r_{+0}\tilde{r}_{-2}-\tilde{r}_{+2}r_{-0})
(\tilde{r}_{+0}r_{-2}-r_{+2}\tilde{r}_{-0})d {\rm log}(a_0-a_2) 
\label{Hamil.0}\\
&+&i r_{+1}r_{-1}d a_1+
i \sum_{j=0,2}(r_{+j}\tilde{r}_{-j}
-\tilde{r}_{j}r_{-j})d a_j -d {\rm log}(a_0-a_2).
\nonumber
\end{eqnarray} 
Here the functions $r_{\pm j}=r_{\pm j}(a_0,a_1,a_2),
~(j=0,1,2),
~\tilde{r}_{\pm 0}=\tilde{r}_{\pm 0}(a_0,a_1,a_2),
~\tilde{r}_{\pm 2}=\tilde{r}_{\pm 2}(a_0,a_1,a_2)$
satisfy the Hamiltonian equations
\begin{eqnarray}
d r_{\pm j}= \{r_{\pm j}, H\},~(j=0,1,2),~~
d \tilde{r}_{\pm 0}= \{\tilde{r}_{\pm 0}, H\}
,~~d \tilde{r}_{\pm 2}= \{\tilde{r}_{\pm 2}, H\}.
\label{Hamil.2}
\end{eqnarray}
where the Poisson bracket is defined by
\begin{eqnarray}
\{r_{+1}, r_{-1}\}=1,~
\{r_{+0}, \tilde{r}_{-0}\}=\{\tilde{r}_{+0}, r_{-0}\}=1,~
\{r_{+2}, \tilde{r}_{-2}\}=\{\tilde{r}_{+2}, r_{-2}\}=1.
\label{Hamil.3}
\end{eqnarray}
This Hamiltonian $H$ depends on
odd number variables $a_0, a_1, a_2$.
In the case without boundary,
the differential equations for
the correlation functions are described by
the Hamiltonian equations which depend on
even number of variables \cite{J.M.M.S.}.
Therefore this point is new for Neumann boundary case.

In the general case, 
we can embed the differential equations for $n$ point
correlation functions of Neumann or Dirichlet boundaries, to the one for $2n$
or $2n-1$ point correlation functions without boundaries.

\begin{thm}~~~In the thermodynamic limit,
the differential equation
for the correlation functions becomes the following.
\begin{eqnarray}
d ~{\rm log}
~\rho_n(x_1',\cdots,x_n'\vert x_1'',\cdots,x_n''\vert 
\varepsilon)
=(1-n)~\omega_{\varepsilon,I_p}\left(\frac{2}{\pi} \right)
+\sum_{j=1}^n\sum_{\sigma \in S_n}
~\omega_{\varepsilon,I_p}^{\left(x_j',x_{\sigma(j)}''\right)}
\left(\frac{2}{\pi}\right),\label{eqn:diff}
\end{eqnarray}
where we denote by $d$ the exterior differentiation with
respect to $x_1',\cdots,x_n', x_1'',\cdots,x_n''$.
Here the differential
forms $\omega_{\varepsilon,I_p}(\lambda)$ and
$\omega_{\varepsilon,I_p}^{(y,y')}(\lambda)$ are defined in
Proposition \ref{omega1} and Proposition \ref{omega2}, respectively.
The differential forms
$\omega_{\varepsilon,I_p}(\lambda)$ and
$\omega_{\varepsilon,I_p}^{(x_j',x_k'')}(\lambda)$ are 
described in terms of solutions of
the generalized fifth Painlev\'e equations
which were introduced by M. Jimbo, T. Miwa, Y. M\^ori and M. Sato
\cite{J.M.M.S.}.
Both Neumann and Dirichlet boundary conditions,
$\omega_{\varepsilon,I_p}(\lambda)$ and
$\omega_{\varepsilon,I_p}^{(y,y')}(\lambda)$ are
described by the same solutions of the same differential equations.
\label{Th:D}
\end{thm}

Physically, it is interesting to derive the long distance
asymptotics of correlation functions :
\begin{eqnarray}
\rho_n(x_1',\cdots,x_n'\vert x_1'',\cdots,x_n''\vert
\varepsilon).
\end{eqnarray}
>From the above theorem,
we can reduce the evaluation of the asymptotics to the following
two step problem.
\begin{enumerate}
\item Evaluate the asymptotics of the solution of the generalized
fifth Painlev\'e introduced in \cite{J.M.M.S.}.
(For our purpose, we only have to consider the special solution
related to the correlation functions for 
the impenetrable Bose gas without boundary. )
\item Determine the asymptotic solutions of the differential equation
(\ref{eqn:diff}) under the appropriate initial condition.
(The main point is to determine the constant multiple in the asymptotics. )
\end{enumerate}

In the case reducible to an ordinary differential equation,
the above two problems have been already solved.
Jimbo, Miwa, M\^ori and Sato \cite{J.M.M.S.}
considered the problem $1$ of the correlation functions
for the impenetrable Bose gas without boundary.
McCoy and Tang \cite{M.T.} 
generalized the asymptotic formulas \cite{J.M.M.S.}
to the $2$-parameter solution of Painlev\'e V,
which is analytic at the origin.
Vaidya and Tracy \cite{V.T.}, \cite{V.}
considered the problem $2$ 
of two-point correlation functions
for the impenetrable
Bose gas without boundary.
(The pioneering work for Ising model was done
by McCoy, Tracy and Wu \cite{M.T.W.}.) 
In our case, to evaluate the asymptotics of
$\rho_1(0 \vert x \vert +)$,
we have to consider the case of three-variables.
However the asymptotics in many variable case is a non-trivial
open problem.
Therefore the above two problems for many variables case 
are our future problems.

\section{Fredholm minor determinant formulas}
The purpose of this section is
to give a proof of Theorem \ref{Th:FD1}.
Set $V={\bf C}^{2}$.
For $\varepsilon=\pm$, define operators 
$\eta^+(\theta,\varepsilon)
, \eta(\theta,\varepsilon)$
acting on $V^{\otimes M}$ by
\begin{eqnarray}
\eta^+(\theta,\varepsilon)
&=&\sum_{m=1}^M
\left(
e^{-i m\theta}+\varepsilon e^{i m\theta}\right)
\psi_m^{+},\\
\eta~(\theta, \varepsilon)&=&\sum_{m=1}^M\left(
e^{i m\theta}+\varepsilon e^{-i m\theta}\right)
\psi_m.
\end{eqnarray}
In the sequel we use the notation
$\theta_{\mu, M}=\frac{\mu}{M+1}\pi$.
The operators
$\eta^{+}(\theta_{\mu,M},\varepsilon),~
\eta(\theta_{\mu,M},\varepsilon)$ have the following
anti-commutation relations for $\varepsilon=\pm,~-M \leq \mu \leq M$.
\begin{eqnarray}
\left\{ \eta^{+}(\theta_{\mu,M}, \varepsilon),
        \eta^{+}(\theta_{\nu,M}, \varepsilon) \right\}&=&
\left\{ \eta~(\theta_{\mu,M}, \varepsilon),
        \eta~(\theta_{\nu,M}, \varepsilon) \right\}=0,\\
\left\{ \eta^{+}(\theta_{\mu,M}, \varepsilon),
        \eta~(\theta_{\nu,M}, \varepsilon) \right\}&=&
2(M+1)(\delta_{\mu,\nu}+\varepsilon \delta_{\mu,-\nu}).
\end{eqnarray}
In the sequel we use the following abbreviations.
\begin{eqnarray}
\vert \Omega_{N,M}(+)\rangle
=\vert \Omega_{N,M}(0,0)\rangle,~
\vert \Omega_{N,M}(-)\rangle
=\vert \Omega_{N,M}(\infty,\infty)\rangle,\\
\langle \Omega_{N,M}(+)\vert
=\langle \Omega_{N,M}(0,0)\vert,~
\langle \Omega_{N,M}(-)\vert
=\langle \Omega_{N,M}(\infty,\infty)\vert.
\end{eqnarray}
Using the operators
$\eta^{+}(\theta, \varepsilon),~
\eta(\theta, \varepsilon)$, we can write
\begin{eqnarray}
\vert \Omega_{N,M}(\varepsilon)\rangle
&=&\sqrt{\frac{1}{(2\varepsilon L)^N}}
\eta^+(\theta_{1,M},\varepsilon)\eta^+(\theta_{2,M},\varepsilon)
\cdots \eta^+(\theta_{N,M},\varepsilon)
\vert \Omega_0 \rangle,\\
\langle \Omega_{N,M}(\varepsilon)\vert
&=&\varepsilon^N\sqrt{\frac{1}{(2\varepsilon L)^N}}
\langle \Omega_0 \vert \eta(\theta_{N,M},\varepsilon)
\cdots \eta(\theta_{2,M},\varepsilon)
\eta(\theta_{1,M},\varepsilon).
\end{eqnarray}
The operators
$\eta^+(\theta_{\mu,M},\varepsilon),~\eta(\theta_{\mu,M},\varepsilon)$
act on the vectors
$\vert \Omega_{N,M}(\varepsilon)\rangle,~
\langle \Omega_{N,M}(\varepsilon)\vert$ as follows.
\begin{eqnarray}
&&{\it For} ~~~\vert \mu \vert \leq N,~~~
\eta^{+}(\theta_{\mu,M},\varepsilon) \vert \Omega_{N,M}(\varepsilon)
\rangle =0,~~
\langle \Omega_{N,M}(\varepsilon) 
\vert \eta~(\theta_{\mu,M},\varepsilon)=0.\\
&&{\it For} ~~~\vert \mu \vert>N,~~~
\langle \Omega_{N,M}(\varepsilon) \vert 
\eta^{+}(\theta_{\mu,M},\varepsilon)=0,~~
\eta~(\theta_{\mu,M},\varepsilon) \vert \Omega_{N,M}
(\varepsilon)\rangle=0.
\end{eqnarray}
Define operators $p_m, q_m (m=1,\cdots,M)$ by
\begin{eqnarray}
p_m=\psi_m^{+}+\psi_m,~q_m=-\psi_m^{+}+\psi_m.
\end{eqnarray}
Set 
\begin{eqnarray}
p(\theta)=\sum_{m=1}^M p_m e^{-i m \theta},
q(\theta)=\sum_{m=1}^M q_m e^{-i m \theta}.
\end{eqnarray}
We have
\begin{eqnarray}
p(\theta)+\varepsilon p(-\theta)&=&
\eta^{+}(\theta,\varepsilon)-\eta(\theta,\varepsilon),\\
~q(\theta)+\varepsilon q(-\theta)&=&
\eta^{+}(\theta,\varepsilon)+\eta(\theta,\varepsilon).
\end{eqnarray}
The following relations hold for $m \geq 1$.
\begin{eqnarray}
\sum_{\mu=-M}^{M+1}p(\theta_{-\mu,M})e^{i m\theta_{\mu,M}}
=0,~~\sum_{\mu=-M}^{M+1}q(\theta_{-\mu,M})e^{i m\theta_{\mu,M}}
=0.
\end{eqnarray}
Therefore, $p_m$ and $q_m$
can be obtained by Fourier transformations.
\begin{eqnarray}
p_m&=&\frac{1}{2(M+1)}
\sum_{\mu=-M}^{M+1}
\left(\eta^{+}(\theta_{\mu,M},\varepsilon)
-\eta~(\theta_{\mu,M},\varepsilon)\right)
e^{i m\theta_{\mu,M}},\\
q_m&=&\frac{1}{2(M+1)}
\sum_{\mu=-M}^{M+1}
\left(\eta^{+}(\theta_{\mu,M},\varepsilon)
+\eta~(\theta_{\mu,M},\varepsilon)\right)
e^{i m\theta_{\mu,M}}.
\end{eqnarray}
We define
\begin{eqnarray}
\langle ~{\wp}~ \rangle_{\varepsilon,N} =
\frac{\langle \Omega_{N,M}(\varepsilon) \vert ~{\wp}~ 
\vert \Omega_{N,M}(\varepsilon) \rangle}
{\langle \Omega_{N,M}(\varepsilon) \vert 
\Omega_{N,M}(\varepsilon) \rangle},
\end{eqnarray}
for ${\wp } \in End\left( \left({\bf C}^2 \right)^{\otimes M}
\right)$.
We call $\langle ~{\wp}~ \rangle_{\varepsilon,N}$
the expectation value of the operator ${\wp}$.
\begin{lem}
~~~~The expectation values of the two products of 
$\eta^{+}(\theta_{\mu,M},\varepsilon)$ and
$\eta~(\theta_{\mu,M},\varepsilon)$
are given by
\begin{eqnarray}
\left(
\begin{array}{cc}
\langle \eta^{+}(\theta_{\mu,M},\varepsilon)
\eta^{+}(\theta_{\nu,M}) \rangle_{\varepsilon,N}&
\langle \eta^{+}(\theta_{\mu,M},\varepsilon)
\eta~(\theta_{\nu,M},\varepsilon) \rangle_{\varepsilon,N}\\
\langle \eta~(\theta_{\mu,M},\varepsilon)
\eta^{+}(\theta_{\nu,M},\varepsilon) \rangle_{\varepsilon,N}&
\langle \eta~(\theta_{\mu,M},\varepsilon)
\eta~(\theta_{\nu,M},\varepsilon) \rangle_{\varepsilon,N}\\
\end{array}
\right)\nonumber \\
=2(M+1)(\delta_{\mu,\nu}+\varepsilon\delta_{\mu,-\nu})
\left(
\begin{array}{cc}
0& \theta_1(N-|\mu|)\\
\theta_2(|\mu|-N)& 0
\end{array}
\right).
\end{eqnarray}
where $-M \leq \mu,\nu \leq M$
and $\theta_1(x),~\theta_2(x)$ are the step function
\begin{eqnarray}
\theta_1(x)=
\left\{
\begin{array}{cc}
1&~~,x \geq 0\\
0&~~,x < 0
\end{array}
\right.
~~~~~
\theta_2(x)=
\left\{
\begin{array}{cc}
1&~~,x > 0\\
0&~~,x \leq 0
\end{array}
\right.
\end{eqnarray}
\end{lem}
\begin{prop}\label{exp.1}
~~~~~The expectation values of the products of
$p_m,~q_m$ are given by
\begin{eqnarray}
\left(
\begin{array}{cc}
\langle ~p_l~p_m~ \rangle_{\varepsilon,N}&
\langle ~p_l~q_m~ \rangle_{\varepsilon,N}\\
\langle ~q_l~p_m~ \rangle_{\varepsilon,N}&
\langle ~q_l~q_m~ \rangle_{\varepsilon,N}\\
\end{array}
\right)
=
\left(
\begin{array}{cc}
\delta_{l,m}& -K_{\varepsilon,l,m}\\
K_{\varepsilon,l,m}& \delta_{l,m}
\end{array}
\right).
\end{eqnarray}
where $l,m=1,2,\cdots,M$ and 
\begin{eqnarray}
K_{\varepsilon,l,m}=\delta_{l,m}-\frac{1}{M+1}
\left\{
\frac{\sin \frac{2N+1}{2(M+1)}(l-m)\pi}
{\sin \frac{1}{2(M+1)}(l-m)\pi}
+\varepsilon\frac{\sin \frac{2N+1}{2(M+1)}(l+m)\pi}
{\sin \frac{1}{2(M+1)}(l+m)\pi}
\right\}.
\end{eqnarray}
\end{prop}
{\sl Proof.}~~~~~~
By direct calculation, we can check the following.
\begin{eqnarray}
\langle~p_l~q_m~\rangle_{\varepsilon,N}
&=&
\frac{1}{(2(M+1))^{~2}}\sum_{\mu=-M}^{M+1}\sum_{\nu=-M}^{M+1}
\langle \left( \eta^{+}-\eta~\right)(\theta_{\mu,M},\varepsilon)
\left( \eta^{+}+\eta~\right)(\theta_{\nu,M},\varepsilon) 
\rangle_{\varepsilon,N}
e^{i l\theta_{\mu,M}}
e^{i m\theta_{\nu,M}}\nonumber \\
&=&
\frac{1}{2(M+1)}\sum_{\mu=-M}^{M+1}\sum_{\nu=-M}^{M+1}
(\delta_{\mu,\nu}+\varepsilon\delta_{\mu,-\nu})
\epsilon_{+}(N-|\nu |)e^{i(l\theta_{\mu,M}+
m\theta_{\nu,M})}\nonumber \\
&=&
-\delta_{l,m}+\frac{1}{M+1}
\left\{
\frac{\sin \frac{2N+1}{2(M+1)}(l-m)\pi}
{\sin \frac{1}{2(M+1)}(l-m)\pi}
+\varepsilon\frac{\sin \frac{2N+1}{2(M+1)}(l+m)\pi}
{\sin \frac{1}{2(M+1)}(l+m)\pi}
\right\}.
\end{eqnarray}
Here $\epsilon_{+}(x)$ denotes the sign function
$
\epsilon_{+}(x)=
\left\{
\begin{array}{cc}
1&,~x \geq 0 ,\\
-1&,~x<0 .
\end{array}\right.
$ 
We have used the relation,
\begin{eqnarray}
\sum_{\mu=-M}^{M+1}\epsilon_{+}( N-|\mu |)e^{i s\theta_{\mu}}
=2\left\{(M+1)~\delta_{s,0}
-\frac{\sin \frac{2N+1}{2(M+1)}\pi s}
{\sin \frac{1}{2(M+1)}\pi s}\right\}.
\end{eqnarray}
\hfill $\Box$

We prepare some notations.
Choose
$0 \leq m_1<\cdots <m_{n}\leq M$ and
$0 \leq m_{n+1}<\cdots<m_{2n} \leq M$.
Let $~m_1'\leq m_2' \leq \cdots \leq  m_{2n}' $ such that
$~m_j'=m_{\sigma(j)}~~(\sigma \in S_{2n})$.
Define the interval $I_{j,M}$ and $I_M$ by
$
I_{j,M}=\{l \in {\bf Z}\vert m_{2j-1}'+1\leq l \leq m_{2j}'\},~
I_M=I_{1,M} \cup I_{2,M} \cup \cdots \cup I_{n,M}.
$
Define $t_m,~t_{I_M} \in {\rm End}
\left( \left({\bf C}^2\right)^{\otimes M} \right)$ by
$
t_m=q_1p_1\cdots q_mp_m,~
t_{I_M}=t_{m_1'}\cdots t_{m_{2n}'}
$.
Define
$R_{\varepsilon,ppI_M}(l,m),~R_{\varepsilon,pqI_M}(l,m)
,~R_{\varepsilon,qpI_M}(l,m)$ and $R_{\varepsilon,qqI_M}(l,m)$
by
\begin{eqnarray}
\left(\begin{array}{cc}
R_{\varepsilon,ppI_M}(l,m)& R_{\varepsilon,pqI_M}(l,m)\\
R_{\varepsilon,qpI_M}(l,m)& R_{\varepsilon,qqI_M}(l,m)
\end{array}\right)
=\frac{1}{\langle ~t_{I_M}~ \rangle_{\varepsilon,N}}
\left(
\begin{array}{cc}
\langle p_l~p_m~t_{I_M} \rangle_{\varepsilon,N}
 & \langle p_l~q_m~t_{I_M} \rangle_{\varepsilon,N} \\
\langle q_l~p_m~t_{I_M} \rangle_{\varepsilon,N}
 & \langle q_l~q_m~t_{I_M} \rangle_{\varepsilon,N}
\end{array}\right),
\end{eqnarray}
where $l,m=1,2,\cdots,M$.
Define the matrix
$K_{\varepsilon,I_M}$ by
$(K_{\varepsilon,I_M})_{j,k \in I_M}
=\langle~q_j~p_k~\rangle_{\varepsilon,N}$.
\begin{lem}\label{exp.2}
~~~~~~The expectation value of $t_{I_M}$ is given by
\begin{eqnarray}
\langle ~t_{I_M}~ \rangle_{\varepsilon,N}
=\det K_{\varepsilon,I_M}.
\end{eqnarray}
For $l,m=1,\cdots,M,$
the following relation holds.
\begin{eqnarray}
R_{\varepsilon,pp{I_M}}(l,m)+R_{\varepsilon,qq{I_M}}(l,m)=0,
~R_{\varepsilon,pq{I_M}}(l,m)+R_{\varepsilon,qp{I_M}}(l,m)=0.
\end{eqnarray}
Furthermore $R_{\varepsilon,pp{I_M}}(l,m), 
R_{\varepsilon,pq{I_M}}(l,m),
R_{\varepsilon,qp{I_M}}(l,m)$ and 
$ R_{\varepsilon,qq{I_M}}(l,m)$
have simple formulas.
For $l,m \in I_M$,~
the following relations hold.
\begin{eqnarray}
\left(\begin{array}{cc}
R_{\varepsilon,ppI_M}(l,m)& R_{\varepsilon,pqI_M}(l,m)\\
R_{\varepsilon,qpI_M}(l,m)& R_{\varepsilon,qqI_M}(l,m)
\end{array}\right)
=
\left(
\begin{array}{cc}
\delta_{l,m} & -(K_{\varepsilon,I_M}^{-1})_{l,m} \\
(K_{\varepsilon,I_M}^{-1})_{l,m}& -\delta_{l,m}
\end{array}\right),
\end{eqnarray}
\end{lem}
{\sl Proof.}
~~~~~From Wick's theorem and
$\langle ~p_l~p_m~\rangle_{\varepsilon,N} =\delta_{l,m}$,
we obtain
$\langle ~p_l~p_m~t_{I_M}\rangle_{\varepsilon,N} =
\langle~t_{I_M}~\rangle_{\varepsilon,N}\delta_{l,m}$.
>From this and Wick's theorem,
we can deduce
\begin{eqnarray}
&&\langle~t_{I_M}~\rangle_{\varepsilon,N}~\delta_{m,m'}
=\langle~p_{m'}~p_{m}~t_{I_M}~\rangle_{\varepsilon,N} \nonumber \\
&=&\langle~p_{m'}~p_{m}~\rangle_{\varepsilon,N}
\langle~t_{I_M}~\rangle_{\varepsilon,N}
+\sum_{\lambda \in I_M}\langle~p_{m'}~q_{\lambda}~\rangle_{\varepsilon,N}
\langle~p_{m}~q_{\lambda}t_{I_M}~\rangle_{\varepsilon,N}
-\sum_{\lambda \in I_M}\langle~p_{m'}~p_{\lambda}~
\rangle_{\varepsilon,N}
\langle~p_{m}~p_{\lambda} t_{I_M}~\rangle_{\varepsilon,N} \nonumber \\
&=&\sum_{\lambda \in I_M}
\langle~p_{m'}~q_{\lambda}~\rangle_{\varepsilon,N}
\langle~p_{m}~q_{\lambda}t_{I_M}~\rangle_{\varepsilon,N}
=-\langle~t_{I_M}~\rangle_{\varepsilon,N}
\sum_{\lambda \in I_M}
R_{\varepsilon,pq I_M}(m,\lambda)(K_{\varepsilon,I_M})_{\lambda, m'}
\end{eqnarray}
\hfill $\Box$

We prepare some notations.
Set
\begin{eqnarray}
K_{\varepsilon,N}(x,x')=\frac{\pi}{2L}\left\{
\frac{\sin\frac{2(n+N)+1}{2L}\pi(x-x')}
     {\sin\frac{1}{2L}\pi(x-x')}
    +\varepsilon
\frac{\sin\frac{2(n+N)+1}{2L}\pi(x+x')}
     {\sin\frac{1}{2L}\pi(x+x')}\right\}.
\end{eqnarray}
Define the integral operator ${\hat K}_{\varepsilon,N,J}$ by
\begin{eqnarray}
(\hat{K}_{\varepsilon,N,J}f)(x)=\int_{J}K_{\varepsilon,N}(x,y)f(y)dy.
\end{eqnarray}
Let us denote by
$\det\left(1-\lambda \hat{K}_{\varepsilon,N,J}\right)$
and
$\det\left(
1-\lambda \hat{K}_{\varepsilon,N,J}
\left|
\begin{array}{ccc}
x_1,&\cdots&,x_n\\
x_1',&\cdots&,x_n'
\end{array}\right.
\right),$
the Fredholm determinant and
the $n-$ th Fredholm minor determinant, respectively.
Namely, we have
\begin{eqnarray}
\det\left(1-\lambda \hat{K}_{\varepsilon,N,J}\right)
=\sum_{l=0}^{\infty}
\frac{(-\lambda)^{l}}{l!}
\int_{J}\cdots \int_{J}
dx_1\cdots dx_l
K_{\varepsilon,N}\left(
\begin{array}{ccc}
x_1,&\cdots&,x_l\\
x_1,&\cdots&,x_l
\end{array}
\right).
\end{eqnarray}
\begin{eqnarray}
\det\left(
1-\lambda \hat{K}_{\varepsilon,N,J}
\left|
\begin{array}{ccc}
x_1,&\cdots&,x_n\\
x_1',&\cdots&,x_n'
\end{array} \right.
\right)
&=&\sum_{l=0}^{\infty}
\frac{(-\lambda)^{l+n}}{l!}
\int_{J}\cdots \int_{J}
dx_{n+1}\cdots dx_{n+l} \nonumber \\
&\times&K_{\varepsilon,N}\left(
\begin{array}{cccccc}
x_1,&\cdots&,x_n,&x_{n+1},&\cdots&,x_{n+l}\\
x_1',&\cdots&,x_n',&x_{n+1},&\cdots&,x_{n+l}
\end{array}
\right),
\end{eqnarray}
where we have used
\begin{eqnarray}
K_{\varepsilon,N}\left(
\begin{array}{ccc}
x_1,&\cdots&,x_l\\
x_1',&\cdots&,x_l'
\end{array}
\right)
=\det_{1 \leq j,k \leq l}\left(
K_{\varepsilon,N}(x_j,x_k')\right).
\end{eqnarray}
Set 
\begin{eqnarray}
R_{\varepsilon,N,J}(x,x'\vert \lambda)
=\sum_{l=0}^{\infty}
\lambda^l
\int_J\cdots \int_J dx_1 \cdots dx_l
K_{\varepsilon,N}(x,x_1)K_{\varepsilon,N}(x_1,x_2)\cdots 
K_{\varepsilon,N}(x_l,x').
\end{eqnarray}
Define the integral operator ${\hat R}_{\varepsilon,N,J}$ by
\begin{eqnarray}
({\hat R}_{\varepsilon,N,J}f)(x)
=\int_{J}R_{\varepsilon,N,J}(x,y\vert \lambda )f(y)dy.
\end{eqnarray}
The resolvent kernel
$R_{\varepsilon,N,J}(x,x'\vert \lambda)$
can be characterized by the following integral equation
\begin{eqnarray}
\left(1-\lambda {\hat K}_{\varepsilon,N,J}\right)
\left(1+\lambda {\hat R}_{\varepsilon,N,J}\right)=1.
\end{eqnarray}
Here we present
a proof of Theorem \ref{Th:FD1}.

{\sl Proof of Theorem \ref{Th:FD1}}
~~~~~~First, for simplicity, we show the $n=1$ case.
For $s_1 \leq s_2, ~(s_1,s_2 \in \{1,2,\cdots,M\})$, we have
\begin{eqnarray}
&&\langle ~\phi_{s_1}~\phi_{s_2}^+~\rangle_{\varepsilon,N}
=\frac{1}{2}
\langle (\phi_{s_1}^++\phi_{s_1})
(\phi_{s_2}^+ +\phi_{s_2})\rangle_{\varepsilon,N}
=\frac{1}{2}
\langle (\psi_{s_1}^+-\psi_{s_1})
\sigma_{s_1+1}^z \cdots \sigma_{s_2-1}^z
(\psi_{s_2}^+ +\psi_{s_2})\rangle_{\varepsilon,N}\nonumber\\
&=&\frac{1}{2}(-1)^{s_2-s_1}
\langle
(q_{s_1}p_{s_1+1})(q_{s_1+1}p_{s_1+2})\cdots
(q_{s_2-1}p_{s_2})\rangle_{\varepsilon,N}
\end{eqnarray}
Applying Wick's theorem and $\langle~p_j~p_k~\rangle_{\varepsilon,N}
=\delta_{j,k}$
and $\langle~q_j~q_k~\rangle_{\varepsilon,N}=\delta_{j,k}$,
we can write the above as a determinant.
\begin{eqnarray}
\langle ~\phi_{s_1}~\phi_{s_2}^+~\rangle_{\varepsilon,N}
=\frac{1}{2}
\det_{s_1\leq j,k \leq s_2-1}\left(
\langle~p_{j+1}~q_k~\rangle_{\varepsilon,N} \right).
\end{eqnarray}
>From (\ref{continuum}),$~
\rho_{1,N,L}(x_1 \vert x_1'\vert \varepsilon)
=\lim_{\epsilon \to 0}
\epsilon^N
\langle~\phi_{s_1}~\phi_{s_2}^+~\rangle_{\varepsilon,1+N}$ holds,
where $\epsilon=\frac{L}{M+1}$ and
$\epsilon s_1 \to x_1,~\epsilon s_2 \to x_1'$.
Set $\nu =s_2-s_1$. From Proposition \ref{exp.1}, we obtain
\begin{eqnarray}
\rho_{1,N,L}(x_1 \vert x_1'\vert \varepsilon)
=\frac{-1}{2(x_1'-x_1)}
\lim_{\nu\to \infty}
\nu \det_{1 \leq j,k \leq \nu}
\left(
-\delta_{j+1,k}+\frac{1}{\nu}G_{\varepsilon}
\left(\frac{j+1+s_1}{\nu}, \frac{k}{\nu}\right)\right),
\end{eqnarray}
where we set
\begin{eqnarray}
G_{\varepsilon}(y,y')
=\frac{x_1'-x_1}{L}\left\{
\frac{\sin\frac{2(1+N)+1}{2L}\pi(y+y')(x_1'-x_1)}
     {\sin\frac{1}{2L}\pi(y+y')(x_1'-x_1)}
    +\varepsilon
\frac{\sin\frac{2(1+N)+1}{2L}\pi(y-y')(x_1'-x_1)}
     {\sin\frac{1}{2L}\pi(y-y')(x_1'-x_1)}\right\}.
\end{eqnarray}
We apply the following relation to the above equation.
\begin{eqnarray}
&&\lim_{\nu \to \infty}
\nu \det_{1 \leq j,k \leq \nu}
\left(-\delta_{j+1,k}+\frac{1}{\nu}\lambda
H\left(\frac{j+1}{\nu}, \frac{k}{\nu}\right)\right) =
-(-\lambda)H(0,1)-(-\lambda)^2
\int_{0}^{1}dy_2 H\left(
\begin{array}{cc}
0,&y_2 \\
1,&y_2
\end{array}\right)-\cdots \nonumber \\
&&-(-\lambda)^{m+1}\frac{1}{m!}
\int_{0}^1\cdots \int_{0}^1
dy_2 \cdots dy_{m+1}
H\left(
\begin{array}{cccc}
0,&y_2,&\cdots&,y_{m+1}\\
1,&y_2,&\cdots&,y_{m+1}
\end{array}\right)-\cdots
\end{eqnarray}
Here $H(y_1,y_2)$ is a continuous function, and we use
\begin{eqnarray}
H\left(\begin{array}{ccc}
y_1,&\cdots &y_m\\
y_1',&\cdots &y_m'
\end{array}\right)=
\det_{1\leq j,k \leq m} \left(H(y_j,y_k')\right).
\end{eqnarray}
We can write down
\begin{eqnarray}
&&\rho_{1,N,L}(x_1 \vert x_1'\vert \varepsilon)
=\left(-\frac{1}{2} \right)\left[
\left(-\frac{2}{\pi}\right)K_{\varepsilon,N}(x_1,x_1')
+\left(-\frac{2}{\pi}\right)^2
\int_{x_1}^{x_1'}dy_2 K_{\varepsilon,N}\left(
\begin{array}{cc}
x_1,&y_2 \\
x_1',&y_2
\end{array}\right)+\cdots \right.\nonumber \\
&&+\left.\left(-\frac{2}{\pi}\right)^{m+1}\frac{1}{m!}
\int_{x_1}^{x_1'}\cdots \int_{x_1}^{x_1'}
dy_2 \cdots dy_{m+1}
K_{\varepsilon,N}\left(
\begin{array}{cccc}
x_1,&y_2,&\cdots&,y_{m+1}\\
x_1',&y_2,&\cdots&,y_{m+1}
\end{array}\right)+\cdots ~~\right].
\end{eqnarray}
Now, we have proved $n=1$ case.
Next we shall prove the general case.
>From Proposition \ref{exp.1} and Lemma \ref{exp.2},
we can deduce,
\begin{eqnarray}
\lim_{M \to \infty}
\langle~t_{I_M}~\rangle_{\varepsilon,n+N}=\det\left.
\left(1-\lambda {\hat K}_{\varepsilon,N,I_p}\right)
\right|_{\lambda=\frac{2}{\pi}}.
\label{eq:1}
\end{eqnarray}
>From Lemma \ref{exp.2}, we see
$
\sum_{l\in I_M}(K_{\varepsilon,I_M})_{m,l}
R_{\varepsilon,qpI_M}(l,m')=\delta_{m,m'}.
$
Comparing this relation to the relation
$
\left(1-\lambda {\hat K}_{\varepsilon,N,I_p}\right)
\left(1+\lambda {\hat R}_{\varepsilon,N,I_p}\right)
=1,
$
we can deduce the following.
\begin{eqnarray}
\lim_{M \to \infty}
\left(\frac{M}{L}\right)
R_{\varepsilon,qpI_M}(m_j,m_k)
=\lambda \left. R_{\varepsilon,N,I_p}(x_j,x_k \vert \lambda)
\right|_{\lambda=\frac{2}{\pi}},
\label{eq:2}
\end{eqnarray}
for $m_j \neq m_k,~\frac{L}{M}m_j \to x_j~(j=1,\cdots,n)$.
Choose
$0 \leq m_1<\cdots <m_{n}\leq M$ and
$0 \leq m_{n+1}<\cdots<m_{2n} \leq M$.
Let $~m_1'\leq m_2' \leq \cdots \leq  m_{2n}' $ such that
$~m_j'=m_{\sigma(j)}~~(\sigma \in S_{2n})$.
Set
$
m_j''=\left\{
\begin{array}{cc}
m_j&,~\sigma(j):~odd ,\\
m_j+1&,~\sigma(j):~even .
\end{array}
\right.
$
>From the parity argument, we obtain
\begin{eqnarray}
\langle t_{m_1'}\cdots\phi_{m_j''}
\cdots \phi_{m_k''}\cdots  t_{m_{2n}'}
\rangle_{\varepsilon,n+N}=0,~
\langle t_{m_1'}\cdots\phi^+_{m_j''}
\cdots \phi^+_{m_k''}\cdots  t_{m_{2n}'}
\rangle_{\varepsilon,n+N}=0.
\label{eq:3}
\end{eqnarray}
The expectation value
$
\langle \phi_{m_1''}\phi_{m_2''}\cdots\phi_{m_{n}''}
\phi^+_{m_{n+1}''}\phi^+_{m_{n+2}''}\cdots  \phi^+_{m_{2n}''}
\rangle_{\varepsilon,n+N}
$
can be written as Pfaffian.
(See p.967 of \cite{S.M.J.4}).
Furthermore, from (\ref{eq:3}),
we can write the expectation value as a determinant.
\begin{eqnarray}
&&\frac{\langle \phi_{m_1''}\phi_{m_2''}\cdots\phi_{m_{n}''}
\phi^+_{m_{n+1}''}\phi^+_{m_{n+2}''}\cdots  \phi^+_{m_{2n}''}
\rangle_{\varepsilon,n+N}}
{\langle ~t_{I_M}~ \rangle_{\varepsilon,n+N}}\nonumber \\
&=&
(-1)^{\frac{1}{2}n(n-1)}
\det_{1 \leq j,k \leq n}
\left( \frac{\langle t_{m_1'}\cdots \phi_{m_j''}\cdots t_{m_{n}''}
t_{m_{n+1}''}\cdots \phi^+_{m_{n+k}''}\cdots t_{m_{2n}'}
\rangle_{\varepsilon,n+N}}
{\langle ~t_{I_M}~ \rangle_{\varepsilon,n+N}}
\right) \nonumber \\
&=&\left(-\frac{1}{2}\right)^{n}
\det_{1 \leq j,k \leq n}
\left(R_{\varepsilon,qpI_M}(m_j'',m_{n+k}'')
\right). \label{eq:4}
\end{eqnarray}
>From the equations (\ref{eq:1}), (\ref{eq:2}), (\ref{eq:4})
, we can deduce
\begin{eqnarray}
&&\lim_{M \to \infty}
\left(\frac{M}{L}\right)^n
\langle \phi_{m_1''}\phi_{m_2''}\cdots\phi_{m_{n}''}
\phi^+_{m_{n+1}''}\phi^+_{m_{n+2}''}\cdots
\phi^+_{m_{2n}''}\rangle_{\varepsilon,n+N}\nonumber \\
&=&(-\lambda)^n
\det\left(1-\lambda {\hat K}_{\varepsilon,N,I_p}\right)
\left.\det_{1 \leq j,k \leq n}
\left( R_{\varepsilon,N,I_p}(x_j,x_k'\vert \lambda) \right)\right|
_{\lambda=\frac{2}{\pi}},
\end{eqnarray}
where $\frac{L}{M}m_j' \to x_j,~\frac{L}{M}m_{n+j}'
\to x_j',~(j=1,\cdots,n)$,
when $M \to \infty$.
Using the Fredholm identity,
\begin{eqnarray}
(-\lambda)^n \det\left(1- \lambda \hat{K}_{\varepsilon,N,I_p}\right)
\det_{1 \leq j,k \leq n}
\left( R_{\varepsilon,N,I_p}(x_j,x_k'\vert \lambda) \right)
=\det\left(1- \lambda \hat{K}_{\varepsilon,N,I_p} \left|
\begin{array}{cccc}
x_1, & x_2, & \cdots & ,x_n\\
x_1',& x_2',& \cdots & ,x_n'
\end{array} \right.\right).
\end{eqnarray}
we can deduce the following.
\begin{eqnarray}
&&\lim_{M \to \infty}
\left(\frac{M}{L}\right)^n
\langle \phi_{m_1''}\phi_{m_2''}\cdots\phi_{m_{n}''}
\phi^+_{m_{n+1}''}\phi^+_{m_{n+2}''}\cdots
\phi^+_{m_{2n}''}\rangle_{\varepsilon,n+N}\nonumber \\
&=&\left(-\frac{1}{2} \right)^n
\det
\left(
1- \frac{2}{\pi} \hat{K}_{\varepsilon,N,I_p} \left|
\begin{array}{cccc}
x_1, & x_2, & \cdots & ,x_n\\
x_1',& x_2',& \cdots & ,x_n'
\end{array} \right.\right)
\end{eqnarray}
This complete the proof of the general case.
\hfill
$\Box$

Fredholm minor series in
this correlation function
is a finite sum because
\begin{eqnarray}
K_{\varepsilon,N}\left(
\begin{array}{ccc}
x_1,&\cdots&,x_l\\
x_1',&\cdots&,x_l'
\end{array}
\right)=0~~~\mbox{for}~~m \geq 2(n+N).\label{eq:5}
\end{eqnarray}
To see this, define an 
$m \times M$ matrix $A_M(\alpha \vert x_1,\cdots,x_m)$ by
\begin{eqnarray}
\left(A_M(\alpha \vert x_1,\cdots,x_m)\right)_{j,k}
=e^{i \alpha k x_j},~~\mbox{for}~~
j=1,\cdots,m,~k=-\frac{1}{2}(M-1),\cdots,\frac{1}{2}(M-1).
\end{eqnarray}
Using this matrix, we obtain the following.
\begin{eqnarray}
&&K_{\varepsilon,N}\left(
\begin{array}{ccc}
x_1,&\cdots&,x_l\\
x_1',&\cdots&,x_l'
\end{array}
\right)=
\sum_{\epsilon_1,\cdots,\epsilon_m=\pm}
(\epsilon_1\cdots \epsilon_m)^{\frac{1-\varepsilon}{2}}
\det_{1 \leq j,k \leq m}
\left(\frac{\pi}{2L}
\frac{\sin\frac{2(n+N)+1}{2L}\pi(x_j-\epsilon_k x_k')}
     {\sin\frac{1}{2L}\pi(x_j-\epsilon_k x_k')}\right)\\
&=&\left(\frac{\pi}{2L}\right)^m
\sum_{\epsilon_1,\cdots,\epsilon_m=\pm}
(\epsilon_1\cdots \epsilon_m)^{\frac{1-\varepsilon}{2}}
\det\left( A_{2(n+N)+1}\left(\left.\frac{\pi}{L}\right|
x_1,\cdots,x_m\right)A^T_{2(n+N)-1}\left(\left.-\frac{\pi}{L}\right|
\epsilon_1x_1',\cdots,\epsilon_mx_m'\right)\right).\nonumber
\end{eqnarray}
Here $A^T$ represents the transposed matrix.
>From elementary argument of linear algebra, we can see
$
\det \left(A_{M}\left(\left.\alpha \right|
x_1,\cdots,x_m\right)A^T_{M}\left(\left.\beta \right|
x_1',\cdots,x_m'\right)\right)=0,
$
for $m \geq M+1$.
Now we have proved (\ref{eq:5}).

\begin{section}{Generalized fifth Painlev\'e equation}
The purpose of this section is to give
a proof of Theorem \ref{Th:D}.
Following \cite{J.M.M.S.}, we describe the correlation functions
in terms of the generalization of the fifth Painlev\'e
equations, which are given by Jimbo, Miwa, M\^ori and Sato
in the thermodynamic limit
$(N,L \to \infty,~\frac{N}{L}=\rho_0:\mbox{fixed}).$
Set
\begin{eqnarray}
K_{\varepsilon}(x,x')=
\frac{\sin\rho_0 \pi (x-x')}
     {x-x'}
    +\varepsilon
\frac{\sin\rho_0 \pi (x+x')}
     {x+x'}.
\end{eqnarray}
Define the integral operators
${\hat K}_{\varepsilon,J}$ by
\begin{eqnarray}
\left({\hat K}_{\varepsilon,J}f\right)(x)
=\int_{J}K_{\varepsilon}(x,y)f(y)~dy.
\end{eqnarray}
Set 
\begin{eqnarray}
R_{\varepsilon,J}(x,x'\vert \lambda)
=\sum_{l=0}^{\infty}\lambda^l
\int_J \cdots \int_J dx_1 \cdots dx_l
K_{\varepsilon}(x,x_1)
K_{\varepsilon}(x_1,x_2)\cdots
K_{\varepsilon}(x_l,x').
\end{eqnarray}
Define the integral operators $\hat{R}_{\varepsilon,J}$ by
\begin{eqnarray}
(\hat{R}_{\varepsilon,J}f)(x)=\int_J
R_{\varepsilon,J}(x,y\vert \lambda)f(y)dy.
\end{eqnarray}
The resolvent kernel $R_{\varepsilon,J}(x,x'\vert \lambda)$
is characterized by the following integral equation,
\begin{eqnarray}
(1+\lambda \hat{R}_{\varepsilon,J})
(1+\lambda \hat{K}_{\varepsilon,J})=1.\label{resol1}
\end{eqnarray}
Let us denote by
$\det\left(1-\lambda \hat{K}_{\varepsilon,J}\right)$
and
$\det\left(
1-\lambda \hat{K}_{\varepsilon,J}
\left|
\begin{array}{ccc}
x_1,&\cdots&,x_n\\
x_1',&\cdots&,x_n'
\end{array}\right.
\right),$
the Fredholm determinant and
the $n-$ th Fredholm minor determinant, respectively.
Namely, we set
\begin{eqnarray}
\det\left(1-\lambda \hat{K}_{\varepsilon,J}\right)
=\sum_{l=0}^{\infty}
\frac{(-\lambda)^{l}}{l!}
\int_{J}\cdots \int_{J}
dx_1\cdots dx_l
K_{\varepsilon}\left(
\begin{array}{ccc}
x_1,&\cdots&,x_l\\
x_1,&\cdots&,x_l
\end{array}
\right).
\end{eqnarray}
\begin{eqnarray}
\det\left(
1-\lambda \hat{K}_{\varepsilon,J}
\left|
\begin{array}{ccc}
x_1,&\cdots&,x_n\\
x_1',&\cdots&,x_n'
\end{array} \right.
\right)
&=&\sum_{l=0}^{\infty}
\frac{(-\lambda)^{l+n}}{l!}
\int_{J}\cdots \int_{J}
dx_{n+1}\cdots dx_{n+l} \nonumber \\
&\times&K_{\varepsilon}\left(
\begin{array}{cccccc}
x_1,&\cdots&,x_n,&x_{n+1},&\cdots&,x_{n+l}\\
x_1',&\cdots&,x_n',&x_{n+1},&\cdots&,x_{n+l}
\end{array}
\right),
\end{eqnarray}
where we have used
\begin{eqnarray}
K_{\varepsilon}\left(
\begin{array}{ccc}
x_1,&\cdots&,x_l\\
x_1',&\cdots&,x_l'
\end{array}
\right)
=\det_{1 \leq j,k \leq l}\left(
K_{\varepsilon}(x_j,x_k')\right).
\end{eqnarray}
We set
\begin{eqnarray}
\rho_{n}(x_1',\cdots ,x_n' \vert x_{1}'',\cdots ,x_{n}''\vert
\varepsilon)
=\lim_{N,L\to \infty, \frac{N}{L}=\rho_0}
\rho_{n,N,L}(x_1',\cdots ,x_n' \vert x_{1}'',\cdots ,x_{n}''
\vert \varepsilon)
\end{eqnarray}

Contrary to the case
in a finite box, the Fredholm minor series in correlation functions is
infinite series and
correlation function $\rho_{n}(x_1',\cdots,x_n'
\vert x_1'',\cdots,x_n''\vert 
\varepsilon)$ becomes a transcendental function.
In the sequel, we shall study the differential equations
for correlation functions in the thermodynamic limit.
In what follows we can choose such a scale
that $\pi \rho_0=1$.

We prepare some notations.
Let $-\infty < a_1 \leq a_2 \leq \cdots
\leq a_{2m}<+\infty.$
We denote by $I$ the interval defined by
$I=[a_1,a_2]\cup \cdots \cup
[a_{2m-1},a_{2m}].$
Set
\begin{eqnarray}
L(x,x')=\frac{\sin (x-x')}{x-x'}.
\end{eqnarray}
Define the integral operators ${\hat L}_{I}$ by
\begin{eqnarray}
(\hat{L}_{I}f)(x)=\int_{I}L(x,y)f(y)dy,
\end{eqnarray}
Set
\begin{eqnarray}
S_{I}(x,x'\vert \lambda)
=\sum_{l=0}^{\infty}
\lambda^l
\int_I \cdots \int_I dx_1 \cdots dx_l
L(x,x_1)L(x_1,x_2)\cdots L(x_l,x').
\end{eqnarray}
Define the integral operator ${\hat S}_{I}$ by
\begin{eqnarray}
({\hat S}_{I}f)(x)=\int_{I}S_{I}(x,y\vert \lambda )f(y)dy.
\end{eqnarray}
The resolvent kernel $S_{I}(x,x'\lambda)$
is characterized by the following integral equation,
\begin{eqnarray}
(1+\lambda\hat{S}_{I})(1-\lambda \hat{L}_{I})=1.
\label{resol2}
\end{eqnarray}
Set
\begin{eqnarray}
S_I\left(\left.\begin{array}{ccc}
x_1,&\cdots,&x_l\\
x_1',&\cdots,&x_l'
\end{array}\right|\lambda \right)
=\det_{1 \leq j,k \leq l}\left(S_I(x_j,x_k'\vert \lambda)\right),
\end{eqnarray}
\begin{eqnarray}
h_{I}(x)=
\frac{1}{2\pi i}{\rm log}
\left\{\frac{(x-a_1)(x-a_3)\cdots (x-a_{2n-1})}
{(x-a_2)(x-a_4)\cdots (x-a_{2n})}
\right\}.
\end{eqnarray}
Set
\begin{eqnarray}
&&S_{I}^{\varepsilon}(x,x'\vert \lambda)
=\sum_{l=0}^{\infty}
\lambda^l
\int_{C_{I}}\cdots
\int_{C_{I}} dy_1 \cdots dy_l\nonumber \\
&\times&
\frac{\varepsilon e^{\varepsilon i(x-y_1)}}
{2i(x-y_1)}h_{I}(y_1)
L(y_1,y_2)h_{I}(y_2)
\cdots L(y_{l-1},y_{l})h_{I}(y_l)
L(y_l,x'),~~~(\varepsilon=\pm),\label{Re.1}
\end{eqnarray}
where the integration
$\oint_{C_{I}}dy_{\mu}$
is along a simple closed curve
$C_{I}$ oriented clockwise, which
encircle the points
$a_1,\cdots,a_{2m}$.
In (\ref{Re.1}), $x$ is supposed to be outside of
$C_{I}$.
We denote $\tilde{S}_{I}^{\varepsilon}
(x,x'\vert \lambda)$
those obtained by letting $x$ inside of
$C_{I}$ in (\ref{Re.1}).
$S_{I}(x,x'\vert \lambda)$
is an entire function in both variables $x,x'$.
$\tilde{S}_{I}^{\varepsilon}
(x,x'\vert \lambda)$
is holomorphic except for a pole at
$x=x'$.
$S_{I}^{\varepsilon}
(x,x'\vert \lambda)$ has branch points at
$x=a_1,\cdots
,a_{2m}$.
The singularity
structure of $S_{I}(x,x'\vert \lambda)$
is as follows.
\begin{eqnarray}
S_{I}^{\varepsilon}
(x,x'\vert \lambda)-
\tilde{S}_{I}^{\varepsilon}
(x,x'\vert \lambda)=\varepsilon
\pi \lambda
h_{I}(x)
S_{I}(x,x'\vert \lambda).
\end{eqnarray}
We set
\begin{eqnarray}
&&S_{\varepsilon,I}(x\vert \lambda)
=\sum_{l=0}^{\infty}
\lambda^l
\int_{C_{I}}\cdots
\int_{C_{I}} dy_1 \cdots dy_l\nonumber \\
&\times&
L(x,y_1)h_{I}(y_1)
L(y_1,y_2)h_{I}(y_2)
\cdots L(y_{l-1},y_l)h_{I}(y_l)
e^{\varepsilon i y_l},~~~(\varepsilon=\pm),\label{Re.2}
\end{eqnarray}
and
\begin{eqnarray}
&&S_{\varepsilon, I}^{~\varepsilon'}
(x\vert \lambda)
=\sum_{l=0}^{\infty}
\lambda^l
\int_{C_{I}}\cdots
\int_{C_{I}} dy_1 \cdots dy_l\nonumber \\
&\times&
\frac{\varepsilon' e^{\varepsilon' i(x-y_1)}}
{2i(x-y_1)}h_{I}(y_1)
L(y_1,y_2)h_{I}(y_2)
\cdots L(y_{l-1},y_l)h_{I}(y_l)
e^{\varepsilon i y_l}
,~~~(\varepsilon, \varepsilon'=\pm),\label{Re.3}
\end{eqnarray}
In (\ref{Re.3}), $x$ is supposed to be outside of
$C_{I}$.
We denote by
$\tilde{S}_{\varepsilon, I}^{~\varepsilon'}
(x\vert \lambda)$
those obtained by letting $x$ inside of
$C_{I}$.
The singularity
structure of $S_{\varepsilon, I}(x,x'\vert \lambda)$
is as follows.
\begin{eqnarray}
S_{\varepsilon, I}^{~\varepsilon'}
(x\vert \lambda)-
\tilde{S}_{\varepsilon, I}^{~\varepsilon'}
(x\vert \lambda)=\varepsilon'
\pi \lambda
h_{I}(x)~
S_{\varepsilon, I}(x\vert \lambda).
\end{eqnarray}
We define the matrices
$Y_I(x),~{\tilde Y}_I(x)$ by
\begin{eqnarray}
Y_I(x)=
\left(
\begin{array}{cc}
S_{+I}(x \vert \lambda) &S_{+I}^{-}(x \vert \lambda) \\
S_{-I}(x \vert \lambda) &S_{-I}^{-}(x \vert \lambda)
\end{array}\right),~~
{\tilde Y}_I(x)=
\left(
\begin{array}{cc}
S_{+I}(x \vert \lambda) &{\tilde S}_{+I}^{-}(x \vert \lambda) \\
S_{-I}(x \vert \lambda) &{\tilde S}_{-I}^{-}(x \vert \lambda)
\end{array}\right).
\label{tilde Y}
\end{eqnarray}
>From the relation (\ref{Re.2}), we obtain
the following monodromy properties.
\begin{eqnarray}
Y_I(x)={\tilde Y}_I(x)
\left\{\frac{(x-a_1)(x-a_3)\cdots (x-a_{2m-1})}
{(x-a_2)(x-a_4)\cdots (x-a_{2m})}\right\}^{
\left(\begin{array}{cc}
0 & \frac{i}{2}\lambda \\
0 & 0
\end{array}
\right)}.
\end{eqnarray}
The matrix $\tilde{Y}_I(x)$ is holomorphic and
$\det \tilde{Y}_I(x)=1.$
It is known
that $Y_I(x)$ satisfies the linear differential equation
(\ref{GP1}). See \cite{J.M.M.S.}.
We define the matrices
$Y_I^{(a,a')}(x)$ and
$\tilde{Y}_I^{(a,a')}(x)$ by
\begin{eqnarray}
Y_I^{(a,a')}(x)
=(x-a)(x-a')\left(\begin{array}{cc}
S_I(x,a\vert \lambda)& S_I^{-}(x,a\vert \lambda)\\
S_I(x,a'\vert \lambda)& S_I^{-}(x,a'\vert \lambda)
\end{array}
\right),\\
\tilde{Y}_I^{(a,a')}(x)
=\left(\begin{array}{cc}
S_I(x,a\vert \lambda)& (x-a)(x-a')
\tilde{S}_I^{-}(x,a\vert \lambda)\\
S_I(x,a'\vert \lambda)& (x-a)(x-a')
\tilde{S}_I^{-}(x,a'\vert \lambda)
\end{array}
\right).
\end{eqnarray}
>From $(\ref{Re.1})$, we obtain the following formula.
\begin{eqnarray}
Y_I^{(a,a')}(x)=
\tilde{Y}_I^{(a,a')}(x)
\left\{(x-a)(x-a')\right\}^
{\left(\begin{array}{cc}
1&0\\
0&0\end{array}\right)}
\left\{\frac{(x-a_1)(x-a_3)\cdots (x-a_{2m-1})}
{(x-a_2)(x-a_4)\cdots (x-a_{2m})}\right\}^{
\left(\begin{array}{cc}
0 & \frac{i}{2}\lambda \\
0 & 0
\end{array}
\right)}
\end{eqnarray}
The matrix $\tilde{Y}_I^{(a,a')}(x)$
is holomorphic and
\begin{eqnarray}
\det \tilde{Y}_I^{(a,a')}(x)
=\frac{a-a'}{2i}S_I(a,a'\vert\lambda).
\end{eqnarray}
Define the matrix $\tilde{Y}_{I\infty}^{(a,a')}(x)$ by
\begin{eqnarray}
\tilde{Y}_{I\infty}^{(a,a')}(x)
=\tilde{Y}_{I}^{(a,a')}(x)
\left(\begin{array}{cc}
1&0\\
-1&1
\end{array}\right)
=(x-a)(x-a')
\left(
\begin{array}{cc}
S_I^{+}(x,a\vert\lambda)&S_I^{-}(x,a\vert\lambda)\\
S_I^{+}(x,a'\vert\lambda)&S_I^{-}(x,a'\vert\lambda)
\end{array}\right).
\end{eqnarray}
The matrix $Y_{I,\infty}^{(a,a')}(x)$
has the following local expansion at $x=\infty$.
\begin{eqnarray}
Y_{I,\infty}^{(a,a')}(x)
=\left(S_{I,\infty}^{(a,a')}+O\left(\frac{1}{x}\right)\right)
~x~ {\rm exp}\left\{x~\left(
\begin{array}{cc}
i&0\\
0&-i
\end{array}\right)\right\},
\end{eqnarray}
where
\begin{eqnarray}
S_{I,\infty}^{(a,a')}=
\left(\begin{array}{cc}
S_{-I}(a \vert \lambda)&S_{+I}(a \vert \lambda)\\
S_{-I}(a' \vert \lambda)&S_{+I}(a' \vert \lambda)
\end{array}\right)
\left(\begin{array}{cc}
-\frac{i}{2}&0\\
0&\frac{i}{2}
\end{array}\right).
\end{eqnarray}
We set
\begin{eqnarray}
Z_I^{(a,a')}(x)={S_{I,\infty}^{(a,a')}}^{-1}
Y_I^{(a,a')}(x),~
\tilde{Z}_I^{(a,a')}(x)={S_{I,\infty}^{(a,a')}}^{-1}
\tilde{Y}_I^{(a,a')}(x).
\label{Z}
\end{eqnarray}
$Z_I^{(a,a')}(x)$ is so normalized that
the local expansion at $x=\infty$ takes the form
\begin{eqnarray}
Z_{I,\infty}^{(a,a')}(x)
=\left(1+O\left(\frac{1}{x}\right)\right)
~x~ {\rm exp}\left\{x~\left(
\begin{array}{cc}
i&0\\
0&-i
\end{array}\right)\right\}.
\end{eqnarray}
Here we start to consider our problem for correlation functions.
Let $0\leq x_1'\leq \cdots \leq x_n '<\infty,~
0\leq x_1''\leq \cdots \leq x_n''<\infty.$
Let $I_p$ the union of $n$ intervals
$I_p=[x_1,x_2]\cup \cdots \cup [x_{2n-1},x_{2n}]$,
where $0\leq x_1\leq \cdots \leq x_{2n}<\infty$
is the re-ordering of
$x_1',\cdots,x_n', x_1'',\cdots ,x_n''.$
Set $I_n=[-x_{2n},-x_{2n-1}]\cup \cdots \cup [-x_{2},-x_1].$
In the sequel, we consider the case
$m=2n,~a_1=-x_{2n},\cdots, a_{2n}=-x_{1}, a_{2n+1}=x_1,\cdots,
a_{4n}=x_{2n}$. We set $I=I_{p}\cup I_{n}.$

\begin{lem}~~~
The resolvent kernel has the following symmetries.
\begin{eqnarray}
&&S_{I_{p}\cup I_{n}}^{\varepsilon}(x,-x'\vert \lambda)
=S_{I_{p}\cup I_{n}}^{-\varepsilon}(-x,x'\vert \lambda),~
\tilde{S}_{I_{p}\cup I_{n}}^{\varepsilon}(x,-x'\vert \lambda)
=\tilde{S}_{I_{p}\cup I_{n}}^{-\varepsilon}(-x,x'\vert \lambda),\\
&&S_{\varepsilon, I_{p}\cup I_{n}}(-x\vert \lambda)
=S_{-\varepsilon, I_{p}\cup I_{n}}(x\vert \lambda),\\
&&S_{\varepsilon,I_{p}\cup I_{n}}^{\varepsilon'}(-x\vert \lambda)
=S_{-\varepsilon,I_{p}\cup I_{n}}^{-\varepsilon'}(x\vert \lambda),~
\tilde{S}_{\varepsilon,I_{p}\cup I_{n}}^{\varepsilon'}
(-x\vert \lambda)
=\tilde{S}_{-\varepsilon,I_{p}\cup I_{n}}^{-\varepsilon'}
(x\vert \lambda).
\end{eqnarray}
\end{lem}
The following is the key lemma.
\begin{lem}~~~The resolvent kernel has
the following linear relation.
\begin{eqnarray}
R_{\varepsilon,I_p}(x,x'\vert \lambda)=
S_{I_{p}\cup I_{n}}(x,x'\vert \lambda)
+\varepsilon S_{I_{p}\cup I_{n}}(x,-x'\vert \lambda)
,~(\varepsilon=\pm).
\label{eqn:key}\end{eqnarray}\label{key}
\end{lem}
{\sl Proof.}~~~~
The following characteristic relation holds,
\begin{eqnarray}
S_{I_{p}\cup I_{n}}(x,x'\vert \lambda)
+\lambda \int_{I_{p}}
\{S_{I_{p}\cup I_{n}}(x,y\vert \lambda)L(y,x')
+S_{I_{p}\cup I_{n}}(x,-y\vert \lambda)L(-y,x')\}dy
=L(x,x').\label{cha}
\end{eqnarray}
>From (\ref{cha}) and the relation $\varepsilon^2=1$,
we derive the following characteristic relation,
\begin{eqnarray}
&&S_{I_{p}\cup I_{n}}(x,x'\vert \lambda)+
\varepsilon S_{I_{p}\cup I_{n}}(x,-x'\vert \lambda) \\
&+&\lambda \int_{I_{p}}(S_{I_{p}\cup I_{n}}(x,y\vert \lambda)
+\varepsilon S_{I_{p}\cup I_{n}}(x,-y\vert \lambda))
(L(y,x')+\varepsilon L(y,-x'))dy
=L(x,x')+\varepsilon L(x,-x').\nonumber
\end{eqnarray}
This means the equation (\ref{eqn:key}).
\hfill $\Box$

Let us derive a formula for
$d~{\rm log} \det\left(1-\lambda \hat{K}_{\varepsilon,I_{p}}\right).$
\begin{prop}~~~~We set 
$\omega_{\varepsilon,I_{p}}(\lambda)
=d~{\rm log} \det\left(1-\lambda 
\hat{K}_{\varepsilon,I_{p}}\right).$
Then we have
\begin{eqnarray}
\omega_{\varepsilon,I_{p}}(\lambda)
&=&trace\left(
\sum_{j=1}^{2n}\tilde{Y}_{I_{p}\cup I_{n}}(x_j)^{-1}
\left.\frac{\partial}{\partial x}\tilde{Y}_{I_{p}\cup I_{n}}(x)
\right|_{x=x_{j}}\left(\begin{array}{cc}
0&\lambda_j\\
0&0
\end{array}\right)dx_j
\right)\nonumber \\
&-\varepsilon&\frac{1}{2} ~trace\left( \sum_{j=1}^{2n}
\tilde{Y}_{I_{p}\cup I_{n}}(x_j)^{-1}
\tilde{Y}_{I_{p}\cup I_{n}}(-x_j)
\left(\begin{array}{cc}
0&\lambda_j\\
0&0
\end{array}\right)\frac{dx_j}{x_j}
\right)\label{tau1.1}\\
&=&trace \left(
\sum_{\delta=\pm}\sum_{1\leq j<k\leq 2n}
\delta \lambda_j \lambda_k
A(x_j)A(\delta x_k)~d~{\rm log} (x_j-\delta x_k)\right)
\nonumber \\
&-\varepsilon&trace \left(
\sum_{j=1}^{2n}\lambda_j
A(x_j)\left\{A_{\infty}dx_j+\frac{1}{2}
\left(\begin{array}{cc}
0&1\\
1&0
\end{array}\right)\frac{dx_j}{x_j}
-\lambda_j\frac{1}{2}A(-x_j)\frac{dx_j}{x_j}\right\} \right).
\label{tau1.2}
\end{eqnarray}
Here the matrix $A(x_j)$ is defined by
\begin{eqnarray}
A(x_j)=\tilde{Y}_{I_{p}\cup I_{n}}(x_j)
\left(
\begin{array}{cc}
0&1\\
0&0
\end{array}
\right)
\tilde{Y}_{I_{p}\cup I_{n}}(x_j)^{-1},~~
\lambda_j=(-1)^{j+1}\frac{i\lambda}{2}.
\end{eqnarray}
\label{omega1}
\end{prop}
{\sl Proof.}
~~~~It is easy to see that
\begin{eqnarray}
\frac{\partial}{\partial x_j}{\rm log}\det\left(
1-\lambda \hat{K}_{\varepsilon,I_{p}}\right)
&=&(-1)^{j+1}\lambda R_{\varepsilon,I_{p}}(x_j, x_j \vert \lambda)\\
&=&(-1)^{j+1}\lambda
\left\{S_{I_{p}\cup I_{n}}(x_j,x_j)
+\varepsilon S_{I_{p}\cup I_{n}}(x_j,-x_j)\right\}.
\end{eqnarray}
>From the definition, we can derive the following formula.
\begin{eqnarray}
\det \left( \begin{array}{cc}
S_{+I}(x) & S_{+I}(x')\\
S_{-I}(x) & S_{-I}(x')
\end{array}\right)=2i(x-x') S_I(x,x').
\end{eqnarray}
Using this formula, we obtain
\begin{eqnarray}
S_I(x,x)&=&\frac{i}{2}
\det\left(\begin{array}{cc}
S_{+I}(x) & \frac{\partial}{\partial x}S_{+I}(x)\\
S_{-I}(x) & \frac{\partial}{\partial x}S_{-I}(x)
\end{array}\right)\\
&=&\frac{i}{2}
trace\left(\tilde{Y}_I(x)^{-1}\frac{\partial}{\partial x}
\tilde{Y}_I(x)
\left(\begin{array}{cc}
0&1\\
0&0
\end{array}\right)\right),\\
S_I(x,-x)&=&\frac{1}{4 ix}
\det\left(\begin{array}{cc}
S_{+I}(x) & S_{+I}(-x)\\
S_{-I}(x) & S_{-I}(-x)
\end{array}\right)\\
&=&\frac{1}{4 i x}
trace\left(\tilde{Y}_I(x)^{-1}\tilde{Y}_I(-x)
\left(\begin{array}{cc}
0&1\\
0&0
\end{array}\right)\right).
\end{eqnarray}
Hence we have the first line
$(\ref{tau1.1})$.
Substituting $(\ref{tilde Y})$ into 
the differential equation,
\begin{eqnarray}
dY_I(x)Y_I(x)^{-1}
=\sum_{j=1}^{2m}
\lambda_j A(a_j) ~d~{\rm log}(x-a_j)
+A_{\infty}dx,\label{GP1}
\end{eqnarray}
 which was
derived in \cite{J.M.M.S.},
and comparing the coefficients of $dx$ at $x=x_j$,
we obtain the following.
\begin{eqnarray}
\left.\frac{\partial}{\partial x}\tilde{Y}_{I_{p}\cup I_{n}}(x)
\right|_{x=x_{j}}\tilde{Y}_{I_{p}\cup I_{n}}(x_j)^{-1}
=
A_{\infty}+
\sum_{\varepsilon=\pm}\sum_{k=1 \atop{k \ne j}}^{2n}
\frac{\varepsilon (-1)^{j+1}}{x_j-\varepsilon x_k}
A(\varepsilon x_k)+\frac{(-1)^{j}}{2x_j}A(-x_j)\nonumber \\
-\tilde{Y}_{I_{p}\cup I_{n}}(x_j)\left\{
\sum_{\varepsilon=\pm}\sum_{k=1 \atop{k \ne j}}^{2n}
\frac{\varepsilon \lambda_k}{x_j-\varepsilon x_k}L_k
-\frac{\lambda_j}{2x_j}L_j\right\}\tilde{Y}_{I_{p}\cup I_{n}}(x_j),
\end{eqnarray}
where $A_{\infty}
=\left(
\begin{array}{cc}
i &0\\
0& -i
\end{array}\right)$.
Substituting this relation into the first line
$(\ref{tau1.1})$, we obtain the second line
$(\ref{tau1.2}).$
\hfill $\Box$

{\sl Remark.}~~~
It is known that the
matrices $A(x_j)$ are solutions of the generalized fifth Painlev\'e
equations introduced in \cite{J.M.M.S.}.

Let us derive a formula for
$d~{\rm log} \det \left(
1-\lambda \hat{K}_{\varepsilon,I_p}\left|
\begin{array}{c}
y\\
y'
\end{array}\right.\right)$.
Let $-\infty <y, y'<+\infty~,(y\ne y')$.
In the sequel, we distinguish the following four cases.
\begin{enumerate}
\item $y, y'\ne x_1, \cdots, x_{2n}.$
\item $y' \ne x_1,\cdots, x_{2n}, y=x_j$ for some $j$.
\item $y \ne x_1,\cdots, x_{2n}, y'=x_{j'}$ for some $j'$.
\item $y=x_j, y'=x_{j'}$ for some distinct $j, j'$.
\end{enumerate}
Set 
\begin{enumerate}
\item $J(y,y')=\left\{0,1,\cdots, 2n+1 \right\}$.
\item $J(y,y')=\left\{0,1,\cdots, 2n+1 \right\}\setminus \left\{ j \right\}$.
\item $J(y,y')=\left\{0,1,\cdots, 2n+1 \right\}\setminus \left\{ j' \right\}$.
\item $J(y,y')=\left\{0,1,\cdots, 2n+1 \right\}
\setminus \left\{ j, j'\right\}$.
\end{enumerate}
Here we set $x_0=y, x_{2n+1}=y'$.
Set $K(y,y')=J(y,y')\setminus \left\{0, 2n+1\right\}$.
Set the notations as follows.
\begin{eqnarray}
M_j=\left(
\begin{array}{cc}
0&1\\
0&0
\end{array}\right),~(j=\pm1,\cdots ,\pm 2n),~M_0=M_{2n+1}=\left(
\begin{array}{cc}
1&0\\
0&0
\end{array}\right).
\end{eqnarray}
\begin{eqnarray}
&&\lambda_j^{(y,y')}
=(-1)^{j+1}\lambda\frac{i}{2}(y-x_j)(y'-x_j),~
\lambda_{-j}^{(y,y')}
=(-1)^{j}\lambda\frac{i}{2}(y+x_j)(y'+x_j)~,
(j=1,\cdots, 2n),\nonumber \\
&&\lambda_0^{(y,y')}=\lambda_{2n+1}^{(y,y')}=1.
\end{eqnarray}
Let us state the Proposition.
\begin{prop}~~~
We set $\omega_{\varepsilon,I_{p}}^{(y,y')}(\lambda)
=d~{\rm log} \det \left(
1-\lambda \hat{K}_{\varepsilon,I_p}\left|
\begin{array}{c}
y\\
y'
\end{array}\right.\right)$.
We denote by $d$ the exterior differentiation
with respect to $x_j~(j \in J(y,y')).$
Then we have
\begin{eqnarray}
\omega_{\varepsilon,I_{p}}^{(y,y')}(\lambda)
=\sum_{\delta=\pm}
\delta \left(1+\varepsilon\frac{y-\delta y'}
{y+\delta y'}\Delta^{(y,\delta y')}\right)^{-1}
\left(
\Omega_1^{(y,\delta y')}
-\varepsilon
\Omega_2^{(y,-\delta y')}-dy -dy'
\right).
\end{eqnarray}
Here we set
\begin{eqnarray}
\Omega_1^{(y,y')}
&=&
trace \left(
\sum_{j \in J(y,y')}\lambda_j^{(y,y')}
\tilde{Z}_{I_{p}\cup I_{n}}^{(y,y')}(x_j)^{-1}
\frac{\partial}{\partial x}\left.
\tilde{Z}_{I_{p}\cup I_{n}}^{(y,y')}(x)\right|_{x=x_j}
M_j dx_j \right)\label{subom.1.1}\\
&=&trace
\left(
\sum_{i,j \in J(y,y') \atop{i<j}}
B_i^{(y,y')} B_j^{(y,y')}~d~{\rm log}(x_i-x_j)\right. \nonumber \\
&+&\left.\sum_{i \in J(y,y')} \sum_{j \in K(y,y')}
B_i^{(y,y')}B_{-j}^{(y,y')}\frac{1}{x_i+x_j}dx_i
+\sum_{i \in J(y,y')}B_i^{(y,y')}A_{\infty} dx_i \right),
\label{subom.1.2}\\
\Omega_2^{(y,y')}&=&
trace
\left(
\sum_{j \in K(y,y')}\frac{i}{2}\lambda_j^{(y,y')}
\tilde{Z}_{I_{p}\cup I_{n}}^{(y,y')}(x_j)^{-1}
\tilde{Z}_{I_{p}\cup I_{n}}^{(y,y')}(-x_j)M_j \frac{dx_j}{x_j}
\right),\label{subom.2.1}\\
&=&
-\Delta^{(y,-y')} \\
&\times&
trace\left\{
\sum_{j \in K(y,y')}\left(
S_{I_{p}\cup I_{n},\infty}^{(-y,y')}B_j^{(-y,y')}
{S_{I_{p}\cup I_{n},\infty}^{(-y,y')}}^{-1}
+S_{I_{p}\cup I_{n},\infty}^{(y,-y')}B_j^{(y,-y')}
{S_{I_{p}\cup I_{n},\infty}^{(y,-y')}}^{-1}\right)M_0 \frac{dx_j}{x_j}
\right\},\label{subom.2.2}\nonumber 
\end{eqnarray}
where
\begin{eqnarray}
\Delta^{(y,y')}
&=&\det \left(
{S_{I_{p}\cup I_{n},\infty}^{(y,-y')}}^{-1}
S_{I_{p}\cup I_{n},\infty}^{(y,y')}\right),\\
B_j^{(y,y')}&=&
\lambda_j^{(y,y')}
\tilde{Z}_{I_{p}\cup I_{n}}(x_j)M_j
\tilde{Z}_{I_{p}\cup I_{n}}(x_j)^{-1},~
(j=\pm 1,\cdots \pm 2n,~0,2n+1).
\end{eqnarray}
\label{omega2}
\end{prop}
{\sl Proof}
~~~It is easy to see that
\begin{eqnarray}
\frac{\partial}{\partial x_j}
{\rm log}\det\left(1-\lambda \hat{K}_{\varepsilon,I_{p}}\left|
\begin{array}{c}
y\\
y'
\end{array}\right. \right)
=(-1)^{j+1}\lambda
\frac{R_{\varepsilon,I_{p}}
\left(\left.\begin{array}{cc}
y&x_j\\
y'&x_j
\end{array}\right|\lambda \right)}
{R_{\varepsilon,I_{p}}(y,y'\vert \lambda)},~(j\ne 0, 2n+1).
\end{eqnarray}
Then lemma \ref{key} and the following imply the
$j th$ part of $(\ref{subom.1.1})$.
\begin{eqnarray}
S_{I_{p}\cup I_{n}}
\left(\left.\begin{array}{cc}
y&x_j\\
y'&x_k
\end{array}\right| \lambda\right)
=\frac{(y-x_j)(y'-x_k)}{(y-y')(x_j-x_k)}
S_{I_{p}\cup I_{n}}
\left(\left.\begin{array}{cc}
y&y'\\
x_j&x_k
\end{array}\right| \lambda\right).
\end{eqnarray}
For $j=0, 2n+1$, the following imply the
$j-th$ part of $(\ref{subom.2.1})$.
\begin{eqnarray}
\frac{\partial}{\partial y}
{\rm log}\det\left(1-\lambda \hat{K}_{\varepsilon,I_{p}}\left|
\begin{array}{c}
y\\
y'
\end{array}\right. \right)
=
\frac{\partial}{\partial y}
{\rm log}
R_{\varepsilon,I_{p}}(y,y'\vert \lambda).
\end{eqnarray}
The second lines
$(\ref{subom.1.2}), (\ref{subom.2.2})$
follows from the first ones
by the same argument as in Theorem
\ref{omega1}.
\hfill $\Box$

{\sl Remarks.}~~~
It is known that the matrices
$B_j^{(y,y')}$ and 
$S_{I_{p}\cup I_{n},\infty}^{(-y,y')}B_j^{(-y,y')}
{S_{I_{p}\cup I_{n},\infty}^{(-y,y')}}^{-1}$
are solutions of the generalized fifth
Painlev\'e equations in \cite{J.M.M.S.}.
For special cases $y=x_i,~y'=x_j$, we have the following formula
$\Delta^{(y,y')}$;
\begin{eqnarray}
\frac{x_i-x_j}{x_i+x_j}
\Delta^{(x_i,x_j)}
=\frac{trace\left(A(x_i)A(-x_j)\right)}
{trace\left(A(x_i)A(x_j)\left(
\begin{array}{cc}
0&1\\
1&0
\end{array}
\right)
\right)}.
\end{eqnarray}

Finally we give a proof of Theorem \ref{Th:D}.

{\sl Proof of Theorem \ref{Th:D}}
~~~~Use the following formula
\begin{eqnarray}
R_{\varepsilon,I_{p}}\left(
\begin{array}{ccc}
y_1&\cdots &y_k\\
y_1'&\cdots & y_k'
\end{array}\right)
=
\frac{\det\left(1-\lambda \hat{K}_{\varepsilon,I_{p}}\left|
\begin{array}{ccc}
y_1&\cdots &y_k\\
y_1'&\cdots & y_k'
\end{array}\right.\right)}
{(-\lambda)^k \det \left(1-\lambda 
\hat{K}_{\varepsilon,I_{p}}\right)}.
\end{eqnarray}
and apply Proposition \ref{omega1} and
Proposition \ref{omega2}.
\hfill $\Box$

For $n=1$ and
$0=x'<x$ case,
because $
R_{[0,x]}(0,x \vert \lambda)
=2S_{[-x,x]}(0,x \vert \lambda)$,
the differential equation becomes simpler form.
\begin{eqnarray}
&&\frac{d}{dx}{\rm log}\rho_1(0 \vert x\vert +)=
\left.\frac{\partial}{\partial y}{\rm log}S_{[-x,x]}(0,y\vert \lambda)\right|
_{y=x}
\nonumber \\
&=&trace\left(\left\{\left(B_0(0,-x.x)
+\frac{1}{2}B_1(0,-x,x)\right)\frac{1}{x}+A_{\infty}\right\}
B_0(0,-x,x)\right)-1.
\end{eqnarray}
Here
$
A_{\infty}=\left(\begin{array}{cc}
i&0\\
0&-i
\end{array}
\right)
$.
The $2\times 2$ matrixes $B_j=B_j(a_0,a_1,a_2),~(j=0,1,2)$
depend on three parameters $a_0,a_1,a_2$
and satisfy the following differential systems
that have the singularities at $y=a_0,a_1,a_2,\infty$.
We denote by $d$ the exterior differentiation with respect
to $y, a_0, a_1, a_2$.
\begin{eqnarray}
dZ_{[a_1,a_2]}^{(a_0,a_2)}(y)
=\left( B_0 d {\rm log}(y-a_0)
+B_1 d {\rm log}(y-a_1)+
B_2 d {\rm log}(y-a_2)+
A_{\infty} dy \right)
Z_{[a_1,a_2]}^{(a_0,a_2)}(y),
\end{eqnarray}
where the $2\times 2$ matrices
$Z_{[b_3,b_4]}^{(b_1,b_2)}(y)$
are defined in $(\ref{Z})$.
The integrability condition
\begin{eqnarray}
d\left(d~Z_{[a_0,a_2]}^{(a_1,a_2)}(y)~
Z_{[a_0,a_2]}^{(a_1,a_2)}(y)^{-1}\right)
=
dZ_{[a_0,a_2]}^{(a_1,a_2)}(y)~
Z_{[a_0,a_2]}^{(a_1,a_2)}(y)^{-1}
\wedge
Z_{[a_0,a_2]}^{(a_1,a_2)}(y)~
Z_{[a_0,a_2]}^{(a_1,a_2)}(y)^{-1},
\end{eqnarray}
gives rise to the following closed differential equation.
\begin{eqnarray}
dB_i=-\sum_{j=0 \atop{j \neq i}}^2
[B_i, B_j]d~{\rm log}(a_i-a_j)-[B_i, A_{\infty}] da_i,~(i=0,1,2).
\label{gSch}\end{eqnarray}
The eigenvalues of $B_0, B_2$ is $(0,1)$.
The eigenvalues of $B_1$ is $(0,0)$.
The diagonal of $B_0+B_1+B_2$ is $(1,1)$.
>From the above matrix properties, we reduce
(\ref{gSch}) to the Hamiltonian equations (\ref{Hamil.0}),
(\ref{Hamil.2}) and (\ref{Hamil.3}) which was introduced in
\cite{J.M.M.S.}.
And we have the equation (\ref{tau}).
\end{section}

~

~

{\sl Acknowledgment}
~~~~I wish to thank Professor T. Miwa , Professor M. Jimbo
and Professor M. Kashiwara for their advices. 
This work is partly supported by the Japan Society for the
Promotion of Science.

~

\end{document}